\documentclass[12pt,a4paper]{article}

\pdfoutput=1

\usepackage{amssymb}
\usepackage{amsthm}

\usepackage{amsmath}
\usepackage{graphicx}
\usepackage{cite}
\usepackage{caption}
\usepackage[labelfont=bf,textfont={sl,bf}]{subfig}
\usepackage{verbatim}

\usepackage{dcolumn}

\newcommand{\bc}{\begin{center}}
\newcommand{\ec}{\end{center}}
\newcommand{\be}{\begin{equation}}
\newcommand{\ee}{\end{equation}}
\newcommand{\bea}{\begin{eqnarray}}
\newcommand{\eea}{\end{eqnarray}}
\newcommand{\ba}{\begin{array}}
\newcommand{\ea}{\end{array}}
\newcommand{\ben}{\begin{enumerate}}
\newcommand{\een}{\end{enumerate}}
\newcommand{\bitem}{\begin{itemize}}
\newcommand{\eitem}{\end{itemize}}

\newcommand{\fr}{\frac}
\newcommand{\beal}{\begin{aligned}}
\newcommand{\eeal}{\end{aligned}}
\newcommand{\bpmatrix}{\begin{pmatrix}}
\newcommand{\epmatrix}{\end{pmatrix}}

\newcommand{\nn}{\nonumber}
\newcommand{\crn}{\nonumber \\}

\newcommand{\al}{\alpha}
\newcommand{\eps}{\epsilon}
\newcommand{\vareps}{\varepsilon}

\newcommand{\si}{\sigma}
\newcommand{\Si}{\Sigma}
\newcommand{\ga}{\gamma}
\newcommand{\Ga}{\Gamma}
\newcommand{\de}{\delta}
\newcommand{\De}{\Delta}
\newcommand{\la}{\lambda}

\newcommand{\dd}{{\rm{d}}}
\newcommand{\hsigma}{{\hat{\sigma}}}

\newcommand{\RE}{\operatorname{Re}}
\newcommand{\REt}{\operatorname{\widetilde{Re}}}
\newcommand{\IM}{\operatorname{Im}}

\newcommand{\cO}{{\cal{O}}}
\newcommand{\cL}{{\cal{L}}}
\newcommand{\cA}{{\cal{A}}}
\newcommand{\cM}{{\mathbf{M}}}

\newcommand{\Bran}{{\rm{Br}}}

\newcommand{\eq}[1]{Eq.~(\ref{#1})}
\newcommand{\bib}[1]{Ref.~\cite{#1}}
\newcommand{\fig}[1]{Fig.~\ref{#1}}
\newcommand{\tab}[1]{Table~\ref{#1}}
\newcommand{\sect}[1]{Section~\ref{#1}}
\newcommand{\ssect}[1]{Subsection~\ref{#1}}

\newcommand{\appen}[1]{Appendix~\ref{#1}}

\newcommand{\gev}{{~\text{GeV}}}
\newcommand{\tev}{{~\text{TeV}}}

\newcommand{\fb}{{~\text{fb}}}

\newcommand{\ltff}{{\texttt{LoopTools/FF}}}
\newcommand{\fcv}{{\texttt{FormCalc-6.0}}}
\newcommand{\fas}{{\texttt{FeynArts}}}
\newcommand{\fav}{{\texttt{FeynArts-3.4}}}
\newcommand{\sloops}{{\texttt{SloopS}}}

\newcommand{\bases}{{\texttt{BASES}}}

\newcommand{\vegas}{{\texttt{VEGAS}}}

\newcommand{\ppWHp}{pp\to W^{-}H^{+}}
\newcommand{\ppWHm}{pp\to W^{+}H^{-}}
\newcommand{\ppWHpm}{pp\to W^{\mp}H^{\pm}}

\newcommand{\ggWHpm}{gg\to W^{\mp}H^{\pm}}

\newcommand{\bbWHpm}{b\bar{b}\to W^{\mp}H^{\pm}}

\newcommand{\bgluWHpb}{bg\to W^{-}H^{+}b}
\newcommand{\bbargluWHpbbar}{\bar{b}g\to W^{-}H^{+}\bar{b}}

\newcommand{\bbargluHtbar}{\bar{b}g\to H^{+}\bar{t}}

\newcommand{\CP}{\text{CP }}

\newcommand{\DRb}{\overline{\text{DR}}}
\newcommand{\OS}{{\text{OS}}}
\newcommand{\MSb}{{\overline{\text{MS}}}}
\newcommand{\DIS}{{\text{DIS}}}
\newcommand{\mbmb}{\overline{m}_b(\overline{m}_b)}
\newcommand{\mbMSb}{m_b^{\MSb}}
\newcommand{\mbDRb}{m_b^{\DRb}}

\newcommand{\salpha}{{\sin{\alpha}}}
\newcommand{\calpha}{{\cos{\alpha}}}
\newcommand{\talpha}{{\tan{\alpha}}}
\newcommand{\sbeta}{{\sin{\beta}}}
\newcommand{\cbeta}{{\cos{\beta}}}
\newcommand{\tbeta}{{\tan{\beta}}}
\newcommand{\cbb}{c_{\beta}}
\newcommand{\sbb}{s_{\beta}}
\newcommand{\tbb}{t_{\beta}}

\newcommand{\ca}{c_{\al}}
\newcommand{\sa}{s_{\al}}
\newcommand{\ta}{t_{\al}}
\newcommand{\hs}{\quad}
\newcommand{\ie}{{\it i.e.}}
\newcommand{\eg}{{\it e.g.}}

\def\slashepi{\epsilon_i\kern -.720em {/}}
\def\slashpi{p_i\kern -.600em {/}}
\def\slashp{p\kern -.550em {/}}

\begin{document}
\begin{titlepage}
\vspace*{0.1cm}\rightline{MPP-2010-151}

\vspace{1mm}
\begin{center}

{\Large{\bf{\boldmath{$W^\mp H^\pm$}} production and CP asymmetry at the LHC}}

\vspace{.5cm}

DAO Thi Nhung, Wolfgang HOLLIK and LE Duc Ninh

\vspace{4mm}

{\it Max-Planck-Institut f\"ur Physik (Werner-Heisenberg-Institut), \\
D-80805 M\"unchen, Germany}

\vspace{10mm}
\abstract{ 
The dominant contributions to $W^\mp H^\pm$ production at the LHC are the tree-level 
$b\bar{b}$ annihilation and the $gg$ fusion. 
We perform for the case of the complex MSSM 
a complete calculation of the NLO EW corrections to the
$b\bar{b}$ annihilation channel and a consistent combination with 
other contributions including the standard and SUSY QCD corrections and the $gg$ fusion,
with resummation of the leading radiative corrections to the bottom-Higgs
couplings and the neutral Higgs-boson propagators.
We observe a large CP-violating asymmetry, arising mainly from the $gg$ channel.
}

\end{center}

\normalsize

\end{titlepage}

\section{Introduction}
The discovery of charged Higgs bosons at the running Large Hadron Collider (LHC) 
would be unambiguous evidence for new physics. Important mechanisms to 
produce $H^{\pm}$ include $gb\to tH^-$; $q\bar{q},gg \to H^+H^-$
and $b\bar{b},gg\to W^\pm H^\mp$ (see \cite{Djouadi:2005gj} for a review and references). 
The first and the last channel 
are of particular interest since they allow to search for 
CP-violating effects at the LHC associated with physics beyond the Standard Model.
Recently, the CP-violating asymmetry for $tH^-/\bar{t}H^+$ production 
has been calculated in~\cite{Christova:2008jv}.

There have been many discussions devoted to the $pp\to  W^\pm H^\mp$ processes in 
the Minimal Supersymmetric Standard Model (MSSM) over 
the last two decades. 
These studies assume that all the soft supersymmetry-breaking parameters are real 
and hence CP violation is absent. 
The two main partonic processes are $b\bar b$ annihilation and the loop-induced $gg$ fusion. 
The first study~\cite{Dicus:1989vf}
computed the tree-level $b\bar b$ contribution and the $gg$ process with third-generation quarks in the 
loops using $m_b=0$ approximation. 
This calculation was then extended for finite $m_b$, thus allowing the investigation of 
the process for arbitrary values of $\tan\beta$ (the ratio $v_2/v_1$ of the vacuum 
expectation values of the two Higgs doublets)~\cite{BarrientosBendezu:1998gd, BarrientosBendezu:1999vd}. 
The inclusion of the squark-loop contribution to the $gg$ channel was done in \cite{BarrientosBendezu:2000tu, Brein:2000cv}. 
The next-to-leading order (NLO) corrections to the $b\bar b$ annihilation are
more complicated and not complete as yet;
the full NLO electroweak (EW) corrections are still missing. 
The Standard Model QCD (SM-QCD) corrections were calculated in~\cite{Hollik:2001hy,Gao:2007wz}, 
the supersymmetric-QCD (SUSY-QCD) corrections in~\cite{Zhao:2005mu,Rauch:2008fy}, 
and the Yukawa part of the electroweak corrections in~\cite{Yang:2000yt}. 
There are also studies on the experimental possibility of observing 
$W^\mp H^\pm$ production at the LHC with subsequent hadronic $H^- \to \bar{t} b$ decay~\cite{Moretti:1998xq} 
and leptonic $H^- \to \tau^- \bar{\nu_\tau} $  decay~\cite{Eriksson:2006yt, Hashemi:2010ce}.
 
The aim of this paper is multifold. First, 
we extend the calculation for $pp\to  W^\pm H^\mp$ to the MSSM with complex 
parameters (complex MSSM, or cMSSM). Second, the full NLO EW corrections to the $b\bar{b}$  
annihilation channel are calculated and consistently combined with 
the other contributions to provide the complete NLO corrections to the $pp\to  W^\pm H^\mp$ processes. 
Third, CP-violating effects arising in the cMSSM are discussed. 
The important issues related to the 
neutral Higgs mixing and large radiative corrections to the bottom-Higgs couplings are also systematically addressed. 

In the cMSSM, new sources of CP violation are 
associated with the phases of soft-breaking parameters and of the Higgsino-mass parameter $\mu$. 
Through loop contributions, CP violation also enters the Higgs sector, which is CP conserving at lowest order 
(see for example \cite{Accomando:2006ga} for more details and references). As a consequence, the $h$, $H$ and 
$A$ neutral Higgs bosons in general mix and form the  mass eigenstates $h_{1,2,3}$ with both CP even and 
odd properties, which can have important impact on many physical observables. 

The bottom-Higgs Yukawa couplings are subject to large quantum corrections in the MSSM. 
We use the usual QCD running bottom-quark mass to absorb large 
QCD corrections to the LO results. The potentially large SUSY-QCD corrections, 
in the large $\tan\beta$ limit, are included into the quantity $\Delta m_b$, 
which is complex in the cMSSM and can be resummed (\sect{running_mb}).

The paper is organized as follows. 
\sect{sect-bbWH} is devoted to the subprocess $b\bar b\to W^\pm H^\mp$, including the issues of 
effective bottom-Higgs couplings and neutral Higgs mixing.
The calculation of the $gg$ fusion part is shown in \sect{sect-ggWH}. Hadronic cross 
sections and CP-violating asymmetry are defined in \sect{sec:hadronic}. Numerical results are presented 
in \sect{sect-results} and conclusions in~\sect{sect-conclusions}.
Feynman diagrams, counterterms, and renormalization constants 
can be found in the Appendices.

\section{The subprocess {$\boldmath{\bbWHpm}$} }
\label{sect-bbWH}
\begin{figure}[h]
  \centering
  \includegraphics[width=0.6\textwidth]{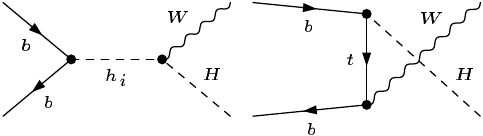}
  \caption{{\em Tree-level diagrams for the partonic process 
$b\bar b\to W^\pm H^\mp$. $h_i$ with $i=1,2,3$ denote the neutral Higgs bosons $h$, $H$ and 
$A$, respectively.}}
  \label{proc_bbWH_born}
\end{figure}
At the tree level, there are four Feynman diagrams including three $s$-channel diagrams with 
a neutral Higgs exchange and a $t$-channel diagram, as shown in \fig{proc_bbWH_born}. 
The tree-level bottom-Higgs couplings read as follows,
\bea
\lambda_{b\bar{b}h}&=&\fr{iem_b}{2s_WM_W}\fr{\salpha}{\cbeta}(P_L + P_R),\crn
\lambda_{b\bar{b}H}&=&\fr{-iem_b}{2s_WM_W}\fr{\calpha}{\cbeta}(P_L + P_R),\crn
\lambda_{b\bar{b}A}&=&\fr{em_b}{2s_WM_W}\tbeta(P_L - P_R),\crn
\lambda_{b\bar{t}H^+}&=&\fr{ie}{\sqrt{2}s_WM_W}\left(\fr{m_t}{\tbeta}P_L + m_b\tbeta P_R\right),\crn
\lambda_{t\bar{b}H^-}&=&\fr{ie}{\sqrt{2}s_WM_W}\left(m_b\tbeta P_L + \fr{m_t}{\tbeta} P_R\right),
\label{b_H_couplings_tree}
\eea
where $P_{L,R}=(1\mp \gamma_5)/2$, $s_W =\sin\theta_W$, and
$\alpha$ is the tree-level mixing angle 
of the two CP-even Higgs bosons.
In order to obtain reliable predictions, 
two important issues related 
to the bottom-Higgs Yukawa couplings and the neutral Higgs mixing have 
to be addressed. These quantities can get large radiative corrections as will be detailed in the next two sections.

\subsection{Bottom-Higgs couplings}
\label{running_mb}
In the context of the MSSM, the bottom-Higgs couplings can get large SM-QCD, SUSY-QCD and EW corrections.  
These large universal corrections can be absorbed into the bottom-Higgs couplings in two steps. 
First, the SM-QCD corrections are absorbed by using the running bottom-quark mass at one-loop order via
\bea
m_b\longrightarrow \mbDRb (\mu_R)=m_b\left[1 - \fr{\alpha_s}{\pi}\left(\fr{5}{3}-\ln\fr{m_b^2}{\mu_R^2}\right)\right].
\eea
We note, in passing, that the relation between the pole mass and the $\MSb$ mass is different
\bea
\mbMSb (\mu_R)=m_b\left[1 - \fr{\alpha_s}{\pi}\left(\fr{4}{3}-\ln\fr{m_b^2}{\mu_R^2}\right)\right].
\eea
It can be proved that by using the running bottom-quark mass in \eq{b_H_couplings_tree} the SM-QCD one-loop 
corrections are independent of $\alpha_s\ln(m_b^2)$ \cite{Braaten:1980yq}.  We will therefore replace 
$m_b=\mbDRb (\mu)$ in \eq{b_H_couplings_tree}. 
$\mbDRb$ can be related to the QCD-$\MSb$ mass $\mbmb$, which is extracted from 
experimental data and is usually taken as 
an input parameter, at two-loop order as follows \cite{Avdeev:1997sz}
\bea
m_b^{\DRb}(\mu_R)=m_b^{\MSb}(\mu_R)\left[1 - \fr{\alpha_s}{3\pi} - \fr{\alpha_s^2}{144\pi^2}(73-3n) \right],
\eea   
where $n$ is the number of active quark flavours and the $\MSb$ running mass is 
evaluated with the two-loop formula  
\begin{eqnarray}
m_b^{\overline{\text{MS}}}(\mu_R) = 
\begin{cases}
U_6(\mu_R, m_t)U_5(m_t, \overline{m}_b)\overline{m}_b(\overline{m}_b) \quad & \text{for}\quad \mu_R > m_t \\
U_5(\mu_R, \overline{m}_b)\overline{m}_b(\overline{m}_b) \quad & \text{for}\quad \mu_R \le m_t
\end{cases}
\label{mb_evolution}
\end{eqnarray}
where the evolution factor $U_n$ reads (see \eg\ \cite{Carena:1999py})
\bea
U_n(Q_2,Q_1)&=&\left(\fr{\alpha_s(Q_2)}{\alpha_s(Q_1)}\right)^{d_n}\left[1 + 
\fr{\alpha_s(Q_1) - \alpha_s(Q_2)}{4\pi}J_n\right],\hs Q_2 > Q_1\crn
d_n&=&\fr{12}{33-2n}, \hs J_n = -\fr{8982 - 504n + 40n^2}{3(33 - 2n)^2}.
\eea

The second step is to absorb large universal SUSY-QCD and EW corrections into the couplings 
in \eq{b_H_couplings_tree}. This is achieved by using the following effective
bottom-Higgs 
couplings \cite{Carena:1999py, Guasch:2003cv, Williams:2008phd, Dittmaier:2009np}:
\bea
\bar\lambda_{b\bar{b}h}&=&\fr{ie\mbDRb}{2s_WM_W}\fr{\salpha}{\cbeta}
\left(\Delta_b^1 P_L + \Delta_b^{1*}P_R\right),\crn
\bar\lambda_{b\bar{b}H}&=&\fr{-ie\mbDRb}{2s_WM_W}\fr{\calpha}{\cbeta}(\Delta_b^{2}P_L + \Delta_b^{2*}P_R),\crn
\bar\lambda_{b\bar{b}A}&=&\fr{e\mbDRb}{2s_WM_W}\tbeta(\Delta_b^{3}P_L - \Delta_b^{3*}P_R),\crn
\bar\lambda_{b\bar{t}H^+}&=&\fr{ie}{\sqrt{2}s_WM_W}\left(\fr{m_t}{\tbeta}P_L + \mbDRb\tbeta \Delta_b^{3*} P_R\right),\crn
\bar\lambda_{t\bar{b}H^-}&=&\fr{ie}{\sqrt{2}s_WM_W}\left(\mbDRb\tbeta \Delta_b^{3} P_L + \fr{m_t}{\tbeta} P_R\right),
\label{b_H_couplings_loop}
\eea  
where
\bea
\Delta_b^1 &=& \fr{1-\Delta_b/(\tbeta\talpha)}{1+\Delta_b},\crn
\Delta_b^2 &=& \fr{1+\Delta_b \talpha/\tbeta}{1+\Delta_b},\crn
\Delta_b^3 &=& \fr{1-\Delta_b/(\tbeta)^2}{1+\Delta_b}.
\eea
The leading corrections 
proportional to $\cO(\alpha_s\tbeta, \alpha_t\tbeta, \alpha\tbeta)$ with 
$\alpha_t=h_t^2/(4\pi)$ and $h_t$ being 
the superpotential top coupling are included 
in $\Delta m_b$ \cite{Carena:1999py}. This quantity is UV finite and can be calculated by 
considering the one-loop corrections to the $H_2^0b\bar{b}$ coupling (which is zero at tree level) 
where $H_2^0$ is the neutral component of the second Higgs doublet. It can also be 
extracted from the one-loop bottom-quark self-energy~\cite{Heinemeyer:2004xw, Hofer:2009xb}. 
In the cMSSM, we find
\bea
\Delta m_b&=&\Delta m_b^{SQCD}+\Delta m_b^{SEW},\crn
\Delta m_b^{SQCD}&=&\fr{2\alpha_s(Q)}{3\pi}M_3^*\mu^*\tan\beta\; I(m_{\tilde{b}_1}^2,m_{\tilde{b}_2}^2,m_{\tilde{g}}^2),\hs 
Q=(m_{\tilde{b}_1} + m_{\tilde{b}_2} + m_{\tilde{g}})/3,\crn
\Delta m_b^{SEW}&=&\Delta m_b^{\tilde{H}\tilde{t}}+\Delta m_b^{\tilde{W}}+\Delta m_b^{\tilde{B}},\crn
\Delta m_b^{\tilde{H}\tilde{t}}&=&\fr{\alpha_t}{4\pi}A_t^*\mu^*\tan\beta\; I(m_{\tilde{t}_1}^2,m_{\tilde{t}_2}^2,\vert\mu\vert^2)\crn
\Delta m_b^{\tilde{W}}&=&-\fr{\alpha}{8\pi s_W^2}M_2^*\mu^*\tbeta \big[ 2\vert U_{11}^{\tilde t}\vert^2 I(m_{\tilde{t}_1}^2,\vert M_2\vert^2,\vert\mu\vert^2) 
+ 2\vert U_{21}^{\tilde t}\vert^2 I(m_{\tilde{t}_2}^2,\vert M_2\vert^2,\vert\mu\vert^2)\crn
& & + \vert U_{11}^{\tilde b}\vert^2 I(m_{\tilde{b}_1}^2,\vert M_2\vert^2,\vert\mu\vert^2)
+ \vert U_{21}^{\tilde b}\vert^2 I(m_{\tilde{b}_2}^2,\vert M_2\vert^2,\vert\mu\vert^2)
\big]\crn
\Delta m_b^{\tilde{B}}&=&-\fr{\alpha}{72\pi c_W^2}M_1^*\mu^*\tbeta 
\big[ 3(\vert U_{11}^{\tilde b}\vert^2 + 2\vert U_{12}^{\tilde b}\vert^2)I(m_{\tilde{b}_1}^2,\vert M_1\vert^2,\vert\mu\vert^2)\crn
& & + 3\, (2\vert U_{22}^{\tilde b}\vert^2 + \vert U_{21}^{\tilde b}\vert^2)I(m_{\tilde{b}_2}^2,\vert M_1\vert^2,\vert\mu\vert^2)
+2I(m_{\tilde{b}_1}^2,m_{\tilde{b}_2}^2,\vert M_1\vert^2)
\big], \label{eq:deltamb}
\eea
with the auxiliary function
\bea
I(a,b,c)=-\fr{1}{(a-b)(b-c)(c-a)}\left(ab\ln\fr{a}{b} + bc\ln\fr{b}{c} 
+ ca\ln\fr{c}{a}\right).
\eea
$M_1$, $M_2$, $M_3$ 
(each with a phase $M_j = \vert M_j\vert e^{i\phi_j}$) and $\mu = \vert\mu\vert
e^{i\phi_\mu}$ are the bino ($\tilde B$), wino ($\tilde W$), 
gluino ($\tilde g$) and Higgsino ($\tilde H$) mass parameters, respectively. 
$A_f = \vert A_f\vert e^{i\phi_f}$, here $f$ means fermion, denotes the soft supersymmetry-breaking trilinear 
scalar coupling. $\tilde b_i$ and $\tilde t_i$ with $i=1,2$ are the 
sbottom and stop mass eigenstates, respectively. 
$U^{\tilde b}$ and $U^{\tilde t}$ are $2\times 2$ mixing matrices.  
By setting all the phases to zero we obtain the results for the real MSSM (rMSSM), 
which agree with those given in \cite{Dittmaier:2006cz, Carena:1999py}.
Since we are also interested in the effect of the $A_b$ phase, corrections proportional 
to $A_b$ are resummed by \cite{Carena:2002bb, Guasch:2003cv}
\bea
\Delta_b&=&\fr{\Delta m_b}{1+\Delta_1},\crn
\Delta_1&=&-\fr{2\alpha_s(Q)}{3\pi}M_3^*A_b I(m_{\tilde{b}_1}^2,m_{\tilde{b}_2}^2,m_{\tilde{g}}^2).
\eea
We remark that $\Delta_b$ is complex and depends on $\phi_{\mu}$, $\phi_{f}$, $\phi_{i}$ with $i=1,2,3$. 
The effective couplings in \eq{b_H_couplings_loop} are used in the calculations of the tree-level, SM-QCD and SUSY-QCD contributions to 
the $\bbWHpm$ process and the $gg$ fusion. For 
the NLO EW corrections we use the tree-level couplings \eq{b_H_couplings_tree} with $m_b = \mbDRb(\mu_R)$.

In the explicit one-loop calculations, we have to subtract the $\Delta_b$-related corrections which 
have already included into the tree-level contribution to avoid double counting. This can 
be done by adding the following counterterms 
\bea
\delta m_b^h&=&\mbDRb\left(1+\fr{1}{\talpha\tbeta}\right)(\Delta_b P_L + \Delta_b^* P_R), \crn
\delta m_b^H&=&\mbDRb\left(1-\fr{\talpha}{\tbeta}\right)(\Delta_b P_L + \Delta_b^* P_R), \crn
\delta m_b^A&=&\mbDRb\left[1+\fr{1}{(\tbeta)^2}\right](\Delta_b P_L - \Delta_b^* P_R),\crn
\delta m_b^{H^+}&=&\mbDRb\left[1+\fr{1}{(\tbeta)^2}\right]\Delta_b^* P_R,\crn  
\delta m_b^{H^-}&=&\mbDRb\left[1+\fr{1}{(\tbeta)^2}\right]\Delta_b P_L
\label{dMB_subtraction}
\eea 
to $\de m_b$ in the corresponding bottom-Higgs-coupling counterterms, 
as listed in  Appendix~\ref{ap:counterterm}. 
Moreover, \eq{dMB_subtraction} is used with $\Delta_b=$ $\Delta m_b^{SQCD}$, 
$\Delta m_b^{SEW}$ for the SUSY-QCD and EW corrections, respectively. 

\subsection{Neutral Higgs-boson propagators}
\label{sect-ggWH-Higgs-propagator}
In the MSSM, the neutral Higgs boson masses are subject to large radiative corrections in 
particular from the Yukawa sector of the theory. As a consequence, the tree-level Higgs masses 
can be quite different from the physical ones. This important effect should be 
considered in the NLO calculations of processes with intermediate neutral Higgs exchange. 

In our calculation, both subprocesses include $s$-channel diagrams with 
internal neutral Higgs bosons, (\fig{proc_bbWH_born} and \fig{proc_ggWH}). 
For $\bbWHpm$ there is also a $t$-channel diagram at tree level which gives the dominant contribution 
at high energies. Thus, the higher-order corrections to the internal Higgs propagators are not 
expected to have important effects in this subprocess at high energies. This will be verified in 
our numerical studies in \sect{sect_results_bbWH_tree}. The situation is different with the 
$gg$ fusion since the $s$-channel (triangle) contribution is large. The higher-order corrections 
to the internal Higgs propagators can therefore be significant in this case, as will be confirmed 
in \sect{sect_results_ggWH}. This issue has not been addressed in the previous studies. 

In a general amplitude with internal neutral Higgs bosons that do not 
appear inside loops, the structure
describing the Higgs-exchange part of an amplitude
is given by
\bea
\cA(p^2) = \sum_{ij} \Ga_i\, \De_{ij}(p^2)\, \Ga_j, \quad i= h,H,A,
\label{amp_higgs_mixing}
\eea 
where $\Ga_{i,j}$ are one-particle irreducible Higgs vertices. 
$p$ is the momentum in the Higgs propagator, which is given in terms  of the
$3\times 3$ propagator matrix
\bea
\De(p^2)& =& i[p^2 -\cM(p^2)]^{-1} , \crn 
 \cM(p^2)& =&
  \bpmatrix
    m_h^2 - \hat{\Si}_{hh}(p^2) & - \hat{\Si}_{hH}(p^2) & - 
\hat{\Si}_{hA}(p^2) \\
    - \hat{\Si}_{hH}(p^2) & m_H^2 - \hat{\Si}_{HH}(p^2) & - 
\hat{\Si}_{HA}(p^2) \\
    - \hat{\Si}_{hA}(p^2) & - \hat{\Si}_{HA}(p^2) & m_A^2 - 
\hat{\Si}_{AA}(p^2)
  \epmatrix  .
\label{eq:propagatormatrix}
\eea
$m_{i}$ ($i=h,H,A$) are the lowest-order Higgs-boson  masses, and $\hat{\Si}_{ij}$ the renormalized
self-energies. 
The physical masses can be found by diagonalizing the above matrix \cite{Frank:2006yh}. 
By using this propagator matrix we effectively resum all the one-loop corrections to the 
neutral Higgs self-energies.

In our calculation, we keep the full propagator matrix and \eq{amp_higgs_mixing} 
whenever neutral Higgs-bosons are exchanged connecting three-point vertices   
in the tree-level $b\bar{b}$ contributions and in the $gg$ fusion diagrams. 
As a consequence, when including the NLO  EW corrections, we 
have to discard all Feynman diagrams containing 
diagonal and nondiagonal $h_ih_j$ self-energies to avoid double counting
(see \fig{proc_bbWH_EW}). 
Whenever neutral Higgs bosons appear inside a loop, 
the tree-level expressions are used for propagators and couplings.

The renormalized Higgs self-energies in \eq{eq:propagatormatrix} are calculated 
at NLO by using the hybrid on-shell and
$\DRb$ scheme (see \sect{sect-bbWHew} and \cite{Frank:2006yh} for details).
Our results have been 
successfully checked against the ones of 
FeynHiggs~\cite{Frank:2006yh,Degrassi:2002fi,Heinemeyer:1998np,Heinemeyer:1998yj}. 
It is noted that FeynHiggs has the option to include the leading two-loop
$\cO(\alpha_s \alpha_t)$ corrections in the cMSSM~\cite{Heinemeyer:2007aq, Hahn:2010te}. 
We have verified that the effects of these two-loop corrections are negligible 
in our numerical analysis and we thus 
chose to perform the numerical evaluation with the one-loop self-energies.

To quantify the effect of the neutral Higgs propagators we introduce
two approximations for the 
subprocess $\bbWHpm$:
The improved-Born approximation (IBA) including both the $\Delta_b$
resummation and the neutral Higgs mixing resummation, and the simpler version
IBA1 which contains only
the resummed $\Delta_b$ together with  tree-level Higgs boson masses and couplings. 
By LO we refer to the tree-level $\bbWHpm$ contribution with $m_b =
m_b^{\DRb}(\mu_R)$ and  the tree-level Higgs sector.

\subsection{SM-QCD corrections}
\label{sect-bbWHsmqcd}

The NLO contribution includes the virtual and real gluonic corrections. 
The virtual corrections, displayed by the Feynman graphs in~\fig{proc_bbWH_SMQCD_virt}, 
contain an extra gluon in the loops.
The calculation is done by using the technique 
of constrained differential renormalisation (CDR) \cite{delAguila:1998nd} which 
is, at one-loop level, equivalent to regularization by dimensional reduction \cite{Siegel:1979wq, Hahn:1998yk}. 
We have also checked by explicit calculations that it is also equivalent to
dimensional regularization \cite{'tHooft:1972fi} 
in this case. 


Concerning renormalization, the bottom-quark mass appearing in the Yukawa couplings is renormalized by using the 
$\DRb$ scheme. It means that the running $m_b^{\DRb}(\mu_R)$ (see \sect{running_mb}) is used in the Yukawa couplings and 
the one-loop counterterm reads
\bea
 \delta m_b^{\DRb} = -m_b\fr{C_F\alpha_s}{4\pi}3C_{UV},
\eea
where  $C_F=4/3$, $C_{UV}=1/\vareps - \gamma_E + \ln(4\pi)$ in $D = 4 - 2\vareps$ space-time dimensions with 
$\gamma_E$ denoting Euler's constant. 
The bottom-quark mass related to the initial state (in the kinematics $p_{b, \bar{b}}^2=m_b^2$ and the spinors) 
is treated as the pole mass since the correct on-shell (OS) behavior must be assured. 
Indeed the $m_b^{\OS}$ effect here is very small and can be neglected. As mentioned in \sect{running_mb},  
the final results are independent of $\ln(m_b^{\OS})$. 
We will therefore set $m_b^{\OS}=m_b^{\DRb}(\mu_R)$ everywhere in this paper. 
The finite wave-function normalization factors for the bottom quarks can be
taken care of by using the OS scheme for 
the wave-function renormalization. 
For the top quark, the  pole mass 
is used throughout this paper. Accordingly, 
the mass counterterm is calculated by using the OS scheme 
(Appendix~\ref{ap:counterterm}). 

The real QCD corrections consist of the  processes with external gluons,
\bea
b + \bar{b} & \to & W^{-} + H^{+} + g,\crn
b + g &\to& b + H^{+} + W^{-},\crn
\bar{b} + g &\to& \bar{b} + W^{-} + H^{+},
\label{pro_NLO_bbWHqcd_real}
\eea
corresponding to the Feynman diagrams shown in \fig{proc_bbWH_realQCD}. 
For the gluon-radiation process, soft and collinear divergences occur.
The soft singularities cancel against those from the virtual corrections, while 
the collinear singularities are regularized by the bottom-quark mass. 
The gluon--bottom-induced processes 
are infrared finite but contain collinear singularities, which are regularized
by the bottom-quark mass as well.
 After adding the virtual and real corrections, the result is collinear divergent and proportional 
to $\ln(m_b^2/\hat{s})$, where $\sqrt{\hat{s}}$ is the center-of-mass energy. 
These singularities are absorbed into 
the bottom and gluon parton distribution functions (PDF), as discussed in~\sect{sec:hadronic}.

Following the line of~\cite{Boudjema:2009pw}, 
we apply both the dipole subtraction scheme~\cite{Catani:1996vz, Dittmaier:1999mb} 
and the two-cutoff phase space slicing method~\cite{Baur:1998kt} to 
extract the singularities from the real corrections. The two techniques give the same results within the 
integration errors. However, the error of the dipole subtraction scheme is
much smaller than the one of the phase space slicing 
method. We will therefore use the dipole subtraction scheme in the numerical analysis.

\subsection{Subtracting the on-shell top-quark contribution}
\label{sect-subtraction-OS}
A special feature of the gluon-induced processes in~(\ref{pro_NLO_bbWHqcd_real})
is the appearance of on-shell top-quarks decaying into $b W$ (and $b H^+$ when
kinematically allowed), which requires a careful treatment and
has been discussed in the previous literature, \eg\ in \cite{Beenakker:1996ch, Tait:1999cf, Frixione:2008yi}. 
Our approach is similar to the one described
in~\cite{Beenakker:1996ch,Tait:1999cf}, with the 
difference that we perform the zero top-quark width limit.

We demonstrate the procedure in terms of
the process $\bbargluWHpbbar$. 
The Feynman diagrams  (\fig{proc_bbWH_realQCD}c)
include a subclass involving the decay $\bar{t}\to \bar{b}W^-$.
When the internal $\bar{t}$ can be on-shell, the propagator 
pole must contain a finite width $\Gamma_t$, which is regarded here as a regulator: 
\bea
\fr{i}{q^2-m_t^2}\longrightarrow \fr{i}{q^2-m_t^2+im_t\Gamma_t}.
\eea
This on-shell contribution is primarily a $\bar{t}H^+$ production and  
should therefore not be considered a NLO contribution.
For the genuine NLO correction, the on-shell top contribution has 
to be discarded in a gauge invariant way. 
Starting from the full set of diagrams, the squared matrix element reads as follows,
\bea
\vert M\vert^2 = \vert M_{\OS}\vert^2 + 2\RE[M_{\OS}M_{\text{non-OS}}^*] 
+ \vert M_{\text{non-OS}}\vert^2,
\label{Msquared}
\eea
where the subscripts $_{\OS}$ and $_{\text{non-OS}}$ denote the contribution of the 
on-shell $\bar{t}$ diagrams and the remainder, respectively. 
The OS part, differential in the $bW$ invariant mass,  
to be subtracted can be identified as 
\bea
\fr{d\sigma^{\bbargluWHpbbar}}{dM_{bW}^2}\bigg\vert_{OS}^{\text{sub}} =
\sigma^{\bbargluHtbar} \Bran(\bar{t}\to \bar{b}W^-) \fr{m_t\Gamma_t }{\pi[(M_{bW}^2-m_t^2)^2+m_t^2\Gamma_t^2]},
\label{bbar_glu_OS1}
\eea
where $\Bran(\bar{t}\to \bar{b}W^-)=\Gamma_{\bar{t}\to \bar{b}W^-}^{LO}/\Gamma_t$. 
The ratio on the right-hand side (rhs) of \eq{bbar_glu_OS1} approaches $\delta(M_{bW}^2-m_t^2)$ when $\Gamma_t\to 0$. 
The subtracted NLO contribution, regularised with the help of $\Gamma_t$, can be
written in the following way,
\bea
\sigma^{\bbargluWHpbbar}_{\text{reg}}(\Gamma_t)&=&\int dM_{bW}^2\left(\fr{d\sigma_{\OS}^{\bbargluWHpbbar}}{dM_{bW}^2}
-\sigma^{\bbargluHtbar}\fr{m_t\Gamma_t\Bran(\bar{t}\to \bar{b}W^-)}{\pi[(M_{bW}^2-m_t^2)^2 + m_t^2\Gamma_t^2]}\right)\crn
&+&\sigma^{\bbargluWHpbbar}_{\text{inter}}
+\sigma^{\bbargluWHpbbar}_{\text{non-OS}},
\label{bbar_gluon_reg1}
\eea 
where the interference and non-OS terms arise from the second and third terms in \eq{Msquared}.
\begin{figure}[t]
  \centering
  \includegraphics[width=0.6\textwidth]{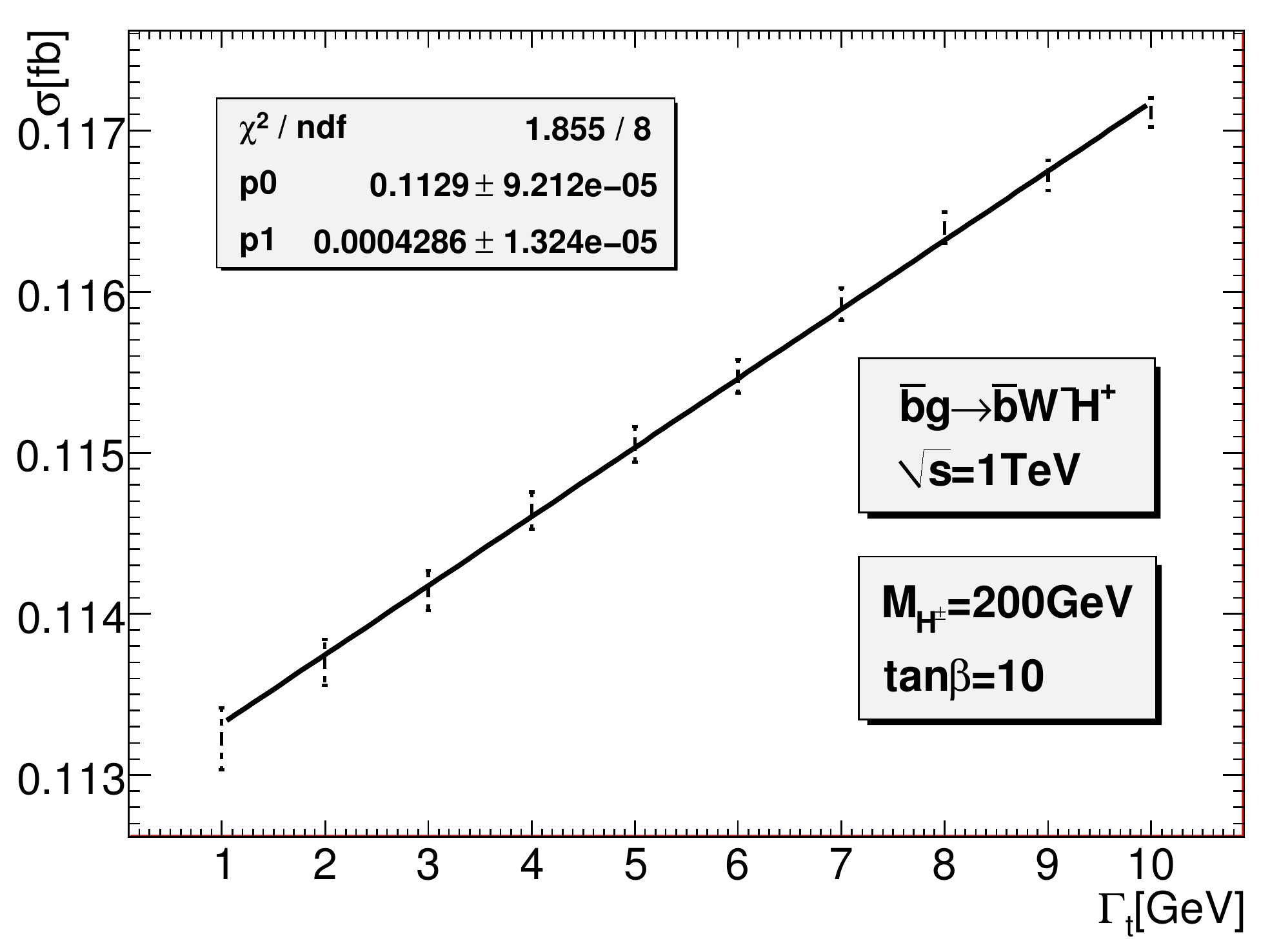}
  \caption{{\em Dependence of the partonic cross section $\sigma^{\bbargluWHpbbar}_{\text{reg}}$ on the width 
  regulator $\Gamma_t$.}}
  \label{Bbarglu_Gamtq_parton}
\end{figure}
There is strong cancellation 
between the first term in the rhs of \eq{bbar_gluon_reg1} and the rest
after subtraction of the collinear part,   
which makes the result of \eq{bbar_gluon_reg1} very small, 
yielding 
an essentially linear dependence on $\Gamma_t$ as displayed in \fig{Bbarglu_Gamtq_parton}. 
We can thus perform the limit $\Gamma_t \to 0$ and obtain a gauge invariant expression by
\bea
\sigma^{\bbargluWHpbbar}_{\text{reg}}=\lim_{\Gamma_t\to 0}\sigma^{\bbargluWHpbbar}_{\text{reg}}(\Gamma_t).
\label{sigma_reg_GamT_lim}
\eea

\begin{figure}[h]
  \centering
  \includegraphics[width=0.6\textwidth]{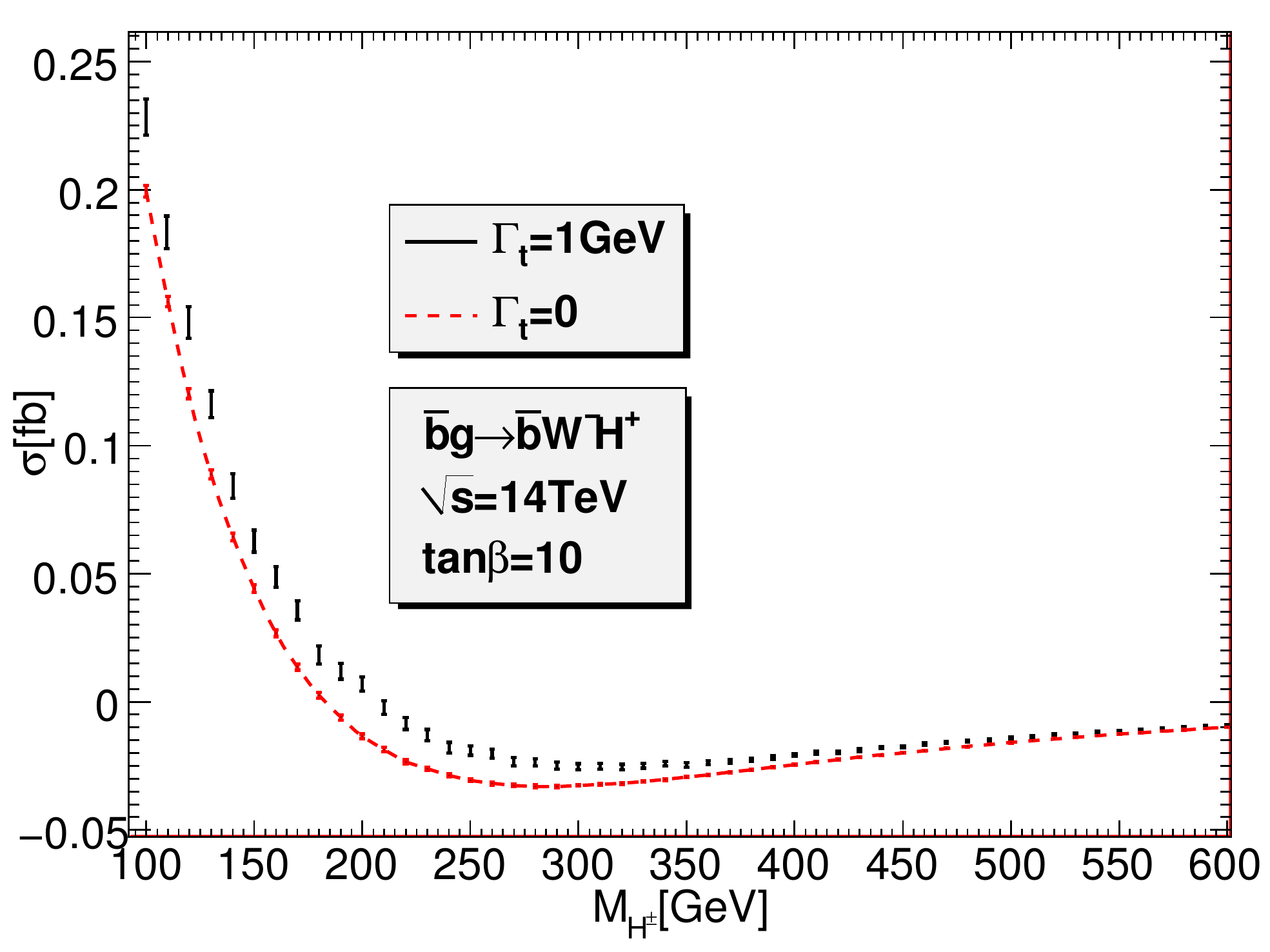}
  \caption{{\em The finite hadronic cross section $\sigma^{\bbargluWHpbbar}_{\text{reg}}$ 
  after subtracting the OS top-quark and the collinear-singularity contributions as 
  a function of $M_{H^\pm}$.}}
  \label{Bbarglu_MHp_hadron}
\end{figure}

\fig{Bbarglu_MHp_hadron} shows that the finite gluon-induced contribution
obtained in this way at the hadronic level (after proper subtraction of the
collinear part)
is very small for large values of $M_{H^\pm}$, but it can be of some 
significance when the charged Higgs boson is light.

The method described above is completely analogous for the process $\bgluWHpb$. 
For low masses, $M_{H^\pm} < m_t$, the 
intermediate on-shell top quark can also decay into $H^+b$. 
This additional OS contribution can be extracted 
by using the same extrapolation method. 
For completeness, we list here the expressions for the decay widths of
 $t\to b W^+$ and $t\to b H^+ $ at lowest order,
\begin{align}
\Ga^{LO}_{t\to b W^+} &= \fr{\al}{16 m_t^3M_W^2s_W^2}(m_t^2-M_W^2)^2(m_t^2+2M_W^2),\\
\Ga^{LO}_{t\to  b H^+} &= \fr{\al}{16 m_t^3M_W^2s_W^2}(m_t^2-M_{H^\pm}^2)^2\left[(\mbDRb\tan\beta)^2
|\De_b^3|^2+\fr{m_t^2}{\tan^2\beta}\right],
\end{align}
 where the $b$-quark mass has been neglected.

\subsection{SUSY-QCD corrections}
The NLO SUSY-QCD contribution consists 
only of the virtual one-loop corrections, visualized  
by the Feynman diagrams with gluino loops in \fig{proc_bbWH_SUSY}. 
The only divergent part is the top-quark self energy, which is renormalized
in the on-shell scheme. 
As discussed 
in \sect{running_mb}, large corrections proportional to $\al_s M_3^* \mu^*\tan\beta$ have been
summed up to all orders in the bottom-Higgs couplings included in the IBA.  
We therefore have to subtract this part
from the explicit one-loop SUSY-QCD corrections to avoid double
counting.

\subsection{Electroweak corrections}
\label{sect-bbWHew}
The full NLO EW contributions to the processes $\bbWHpm$ in the cMSSM have not been computed yet. 
They comprise both virtual and real corrections. 
For the virtual part,  \fig{proc_bbWH_EW} illustrates the various 
classes of one-loop Feynman diagrams. 
As before, the calculation is performed using the CDR technique.
We have also worked out all the 
necessary counterterms in the cMSSM and implemented them in \fav \cite{Hahn:2000kx, Hahn:2001rv}.
Explicit expressions for the counterterms 
can be found in \appen{ap:counterterm}.
For the 
Higgs field renormalization
and $\tan\beta$, we use the $\DRb$ renormalization scheme as specified in \cite{Frank:2006yh}.
Hence, the correct OS behavior of the external $H^\pm$ must be ensured by
including the finite wave-function 
renormalization factor \cite{Hollik:2010dh}
\bea
\sqrt{Z_{H^-H^+}}=1 - \fr{1}{2}\RE\fr{\partial}{\partial p^2}\hat\Sigma_{H^-H^+}(p^2)\big\vert_{p^2=M_{H^\pm}^2},
\eea 
where $\hat\Sigma_{H^-H^+}(p^2)$ is the $H^\pm$ renormalized self-energy. 
The other renormalization constants are determined according to the OS scheme.
To make the EW corrections independent of $\ln m_f$ from the light fermions $f\neq t$, 
we use the fine-structure constant at $M_Z$, $\alpha = \alpha(M_Z)$ as an input parameter. 
This means that we have to modify the counterterm as
\bea
\delta Z_e^{\alpha(M_Z)}&=&\delta Z_e^{\alpha(0)} - \fr{1}{2}\Delta\alpha(M_Z^2),\crn
\Delta\alpha(M_Z^2)&=&\fr{\partial \Sigma_T^{AA}}{\partial k^2}\bigg\vert_{k^2=0}-\fr{\RE\Sigma_T^{AA}(M_Z^2)}{M_Z^2},
\eea
where the photon self-energy includes only the light fermion contribution, to
avoid double counting.

The real EW contributions correspond to the processes with external photons,
\bea
b + \bar{b} &\to& W^{-} + H^{+} + \gamma,\crn
b + \gamma &\to& b + H^{+} + W^{-},\crn
\bar{b} + \gamma &\to& \bar{b} + W^{-} + H^{+},
\label{pro_NLO_bbWHew_real}
\eea
described by the Feynman diagrams of~\fig{proc_bbWH_realEW}. 
They are calculated in the same way as the real QCD corrections, 
discussed in \sect{sect-bbWHsmqcd} and \sect{sect-subtraction-OS}. 
Naively, we would expect this photon contribution to be much smaller than
the one from the gluon, due to the smallness of the 
EW coupling $\alpha$ and the photon PDF. 
This is not always true, however, since the
photon couples to the $W^\pm$ and $H^\pm$ as well. 
The soft singularities are completely cancelled, as in the case of QCD. 
The EW splitting $\gamma\to H^+ H^-$ 
(similarly for $\gamma \to W^+ W^-$), on the other side, 
can introduce large collinear correction in the limit $M_{H^\pm}/Q\to 0$, $Q$ is a typical energy scale. 
The constraint $M_H^\pm > M_W$ prevents those splittings from becoming 
divergent. We observe, however, 
that the finite corrections (after subtracting the collinear bottom-photon and the OS top-quark contributions) 
from the above $\bar{b}\gamma$ process are still larger than the corresponding QCD ones for $M_{H^\pm}<200\gev$, 
\eg\ for $M_H=150\gev$ and $\sqrt{s}=14\tev$ by a factor of 2. 
The photon-induced contribution should thus be included in the NLO
calculations for $W^\pm/H^\pm$ production at high energies. 
This requires the knowledge of the photon density in the proton, which at present
is contained in the set MRST2004qed~\cite{Martin:2004dh} 
of PDFs.

\section{The subprocess {\boldmath{$\ggWHpm$}}}
\label{sect-ggWH}
The subprocess $\ggWHpm$ is loop induced, in the MSSM with quark- and squark-loop contributions. 
\fig{proc_ggWH} summarizes the various one-loop Feynman diagrams, which
involve three- and four-point vertex functions.  
Since the (s)quarks are always coupled to a Higgs boson, the one-loop 
amplitude is proportional to (s)quark-Higgs couplings. The dominant contributions therefore arise
from the diagrams with the third-generation (s)quarks.  
As in \cite{Brein:2000cv},
the contribution from the first two generations of (s)quarks is neglected in this paper. 
Compared to the previous work~\cite{Brein:2000cv}, our calculation is improved by using 
the effective bottom-Higgs couplings and the resummed neutral Higgs propagators. 
It turns out that these improvements sizably affect both the cross section and CP-violating
asymmetry. We have checked our results against those of~\cite{Brein:2000cv}
for the case of the real MSSM using the tree-level couplings and Higgs propagators 
and found good agreement.

\begin{figure}[h]
  \centering
  \includegraphics[width=0.7\textwidth]{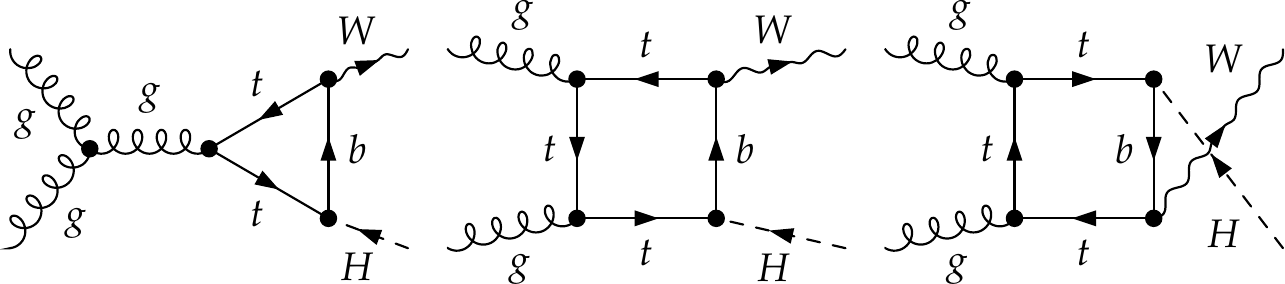}
  \caption{{\em Feynman diagrams that can produce three-point Landau singularities.}}
  \label{fig:quark_singularity}
\end{figure}
We notice an interesting feature related to the anomalous thresholds.
Fig.~1b of \cite{Brein:2000cv} shows a very sharp peak close to the normal $t\bar{t}$ threshold. 
Careful observation reveals that the peak position is slightly above 
$2m_t$ and is 
obviously more singular than the normal thresholds in Fig.~1a of \cite{Brein:2000cv}. This is indeed 
an anomalous threshold corresponding to the three-point Landau singularity 
(see \cite{ninh_bbH2, ninh_thesis} and references therein) of the triangle and box diagrams in 
\fig{fig:quark_singularity}. A simple calculation following \cite{ninh_bbH2} yields the peak position
at
\bea 
\hat{s}_{\text{peak}} &=& \fr{1}{2m_b^2}\big[(M_{H^\pm}^2+M_W^2)(m_t^2+m_b^2)-(m_b^2-m_t^2)^2-
M_{H^\pm}^2M_W^2 \crn&& - \la^{1/2}(m_t^2,m_b^2,M_{H_{\pm}}^2)
\la^{1/2}(m_t^2,m_b^2,M_{W}^2)\big],
\label{eq:landau_s_peak}
\eea
with $\lambda(x,y,z)=x^2+y^2+z^2-2(xy+yz+xz)$. 
The partonic cross section is divergent at $\hat{s}=\hat{s}_{\text{peak}}$ but 
the result is finite at the hadronic level, \ie\ after integrating over
$\hat{s}$, since this singularity is logarithmic and thus integrable. 
The conditions for this anomalous threshold to be in the physical region can also be given~\cite{ninh_bbH2},
\bea 
 2m_t &\le& \sqrt{\hat{s}} \le \sqrt{\fr{m_t}{m_b}[(m_t+m_b)^2-M_W^2]},\crn
 m_b+m_t &\le& M_{H^\pm} \le \sqrt{2(m_t^2+m_b^2) -M_W^2}.
\eea
Similarly, the three-point Landau singularities can occur in the squark diagrams.

\section{Hadronic cross section and CP asymmetry}
\label{sec:hadronic}
The LO hadronic cross section, in terms of the LO partonic $b\bar{b}$
annihilation cross section,
is given by
\bea
\sigma^{pp}_{LO}=
\int \dd x_1\dd x_2[F_b^{p}(x_1, \mu_F)F_{\bar{b}}^{p}(x_2, \mu_F)\hsigma^{b\bar{b}}_{LO}(\alpha^2,\mu_R)+(1\leftrightarrow 2)],
\eea
where $F_{b/\bar{b}}^p(x,\mu_F)$ is the bottom PDF
at momentum faction $x$ and  factorization scale $\mu_F$. 
Other $q\bar{q}$-subprocesses ($q=u,d,c,s$) are neglected due to the smallness of 
light-quark-Higgs couplings. 

The NLO hadronic cross section reads as follows,
\bea
\sigma^{pp}_{NLO}&=&\sum_{i,j}\fr{1}{1+\delta_{ij}}\int \dd x_1\dd x_2[F_i^{p}(x_1, \mu_F)F_j^{p}(x_2, \mu_F)\hsigma^{ij}_{NLO}(\alpha^2,\alpha^2\alpha_s,\alpha^3,\alpha^2\alpha_s^2,\mu_R)\crn
&+&(1\leftrightarrow 2)],
\eea
where  $i,j$ = ($b, \bar{b}, g,\gamma$)  and
\bea
\hsigma^{ij}_{NLO}&=&\hsigma^{b\bar{b}}_{IBA}(\alpha^2)+\Delta_{\text{SM-QCD}}\hsigma^{ij}_{NLO}(\alpha^2\alpha_s)
+\Delta_{\text{SUSY-QCD}}\hsigma^{ij}_{NLO}(\alpha^2\alpha_s)\crn
&+&\Delta_{EW}\hsigma^{ij}_{NLO}(\alpha^3)+\hsigma^{gg}(\alpha^2\alpha_s^2)
\eea
contain the various NLO contributions at the parton level, discussed in the previous sections. 
As mentioned there, the mass singularities of the type $\alpha_s\ln(m_b)$ and
$\alpha\ln(m_b)$ are absorbed in the quark distributions. 
We use the MRST2004qed set of PDFs~\cite{Martin:2004dh}, 
which include $\cO(\alpha_s)$ QCD and $\cO(\alpha)$ photonic corrections. 
As explained in \cite{Diener:2005me}, the consistent
use of these PDFs requires the $\MSb$ factorization scheme for the QCD, but the DIS scheme for
the photonic corrections. We therefore redefine the (anti-)bottom PDF as
follows, 
\bea 
q(x)& =& q(x, \mu_{\text F}^2) -\fr{\al_s C_F}{2\pi}
\int_x^1\fr{dz}{z} q\left(\fr xz, \mu_F^2\right)
 \bigg\{ \ln\bpmatrix\fr{\mu_F^2}{m_b^2}\epmatrix
[P_{qq}(z)]_+\crn
&& - \,[P_{qq}(z)(\ln(1-z)^2 +1)]_+ + C_{qq}^{\MSb}(z) \bigg\} \crn
&& -\fr{\al Q_b^2}{2\pi}
\int_x^1\fr{dz}{z} q\left(\fr xz, \mu_F^2\right)
 \bigg\{ \ln\bpmatrix\fr{\mu_F^2}{m_b^2}\epmatrix
[P_{qq}(z)]_+\crn
&& - \,[P_{qq}(z)(\ln(1-z)^2 +1)]_+ + C_{qq}^{\DIS}(z) \bigg\} \crn
&&-\, \fr{\al_sT_F }{2\pi}\int_x^1\fr{dz}{z} g\left(\fr xz, \mu_F^2\right)
\bigg[ \ln\bpmatrix \fr{\mu_F^2}{m_b^2}\epmatrix P_{qg} + C_{qg}^{\MSb}(z) \bigg]\crn
&&-\, \fr{3\al Q_b^2}{2\pi}\int_x^1\fr{dz}{z} \gamma\left(\fr xz, \mu_F^2\right)
\bigg[ \ln\bpmatrix \fr{\mu_F^2}{m_b^2}\epmatrix P_{q\gamma} + C_{q\gamma}^{\DIS}(z) \bigg]
,
\label{pdf_redifined}
\eea
with $C_F = 4/3$, $T_F=1/2$.
The splitting functions are given by
\bea P_{qq}(z) = \fr{1+z^2}{1-z}, \quad
P_{qg}(z) = P_{q\gamma}(z) = z^2 + (1-z)^2,\eea 
and the $[\ldots]_{+}$ prescription is understood in the usual way,
\bea
\int_x^1dzf(z)\left[\fr{g(z)}{1-z}\right]_{+}=\int_x^1dz\fr{[f(z)-f(1)]g(z)}{1-z}-f(1)\int_0^xdz\fr{g(z)}{1-z}.
\eea
Following the standard conventions of QCD, the factorization schemes are
specified by 
\bea
C_{qq}^{\MSb}(z) &=& C_{qg}^{\MSb}(z) = 0,\crn
C_{qq}^{\DIS}(z) &=& \left[P_{qq}(z)\left(\ln(\fr{1-z}{z}) - \fr{3}{4}\right) + \fr{9+5z}{4} \right]_+ ,\crn
C_{q\gamma}^{\DIS}(z) &=& P_{q\gamma}\ln(\fr{1-z}{z}) - 8z^2 + 8z - 1.
\eea
Having constructed in this way the hadronic cross sections $\sigma(pp\to W^\pm
H^\mp)$, we can define
the CP-violating asymmetry at the hadronic level in the following way,
\bea
\delta^{\CP}_{pp}&=&\fr{\sigma(\ppWHp)-\sigma(\ppWHm)}{\sigma(\ppWHp)+\sigma(\ppWHm)}.
\label{eq_delta_cp_pp}
\eea
The numerator gets contributions from the NLO-$b\bar{b}$ corrections (the LO is CP conserving) and the loop-induced $gg$ process. 
However, the latter is much larger than the former due to the dominant gluon PDF. The CP-violating effect is 
therefore mainly generated by the $gg$ channel. The LO-$b\bar{b}$ contribution adds only to the 
CP invariant part and therefore reduces the magnitude of the CP asymmetry.

\section{Numerical studies}
\label{sect-results}

\subsection{Input parameters}
\label{sect-input}
We use the following set of input parameters for the SM sector \cite{Amsler:2008zzb,
:2009ec}, 
\begin{equation}
\begin{aligned}
\alpha_s(M_Z) &= 0.1197, \hs &\alpha(M_Z)&=1/128.926, \\
M_{W} &= 80.398\gev, \hs& M_Z&= 91.1876\gev, \\
m_t & =173.1\gev, \hs &\mbmb& = 4.2\gev.
\end{aligned}
\end{equation} 
We take here $\alpha_s = \alpha_s^{\MSb}(\mu_R)$ at three-loop order \cite{Amsler:2008zzb}.
 $\mbmb$ is the QCD-$\MSb$ $b$-quark mass, while the top-quark mass is understood 
as the pole mass. 
CKM matrix elements are approximated by $V_{td}=V_{ts}=0$ and $V_{tb}=1$. 

For the soft SUSY-breaking parameters, we use the adapted CP-violating benchmark scenario (CPX)~\cite{Williams:2007dc,Carena:2000ks},
\bea\beal 
|\mu| &= 2\tev, |M_2|=200\gev,\, |M_3| = 1\tev,\, |A_t|=|A_b|=|A_\tau|=900\gev,\\
M_{\tilde Q}&=M_{\tilde D}=M_{\tilde U}=M_{\tilde L}=M_{\tilde
  E}=M_{\text{SUSY}}=500\gev .
\eeal\eea
Since the Yukawa couplings of the first two fermion generations proportional to 
the small fermion masses are neglected in our calculations, we set $A_f=0$ for 
$f=e,\mu,u,d,c,s$. The values of $M_1$ and $M_2$ are connected via the GUT relation
 $|M_1|= 5/3\tan^2\theta_W |M_2|$.   
We can set $\phi_2 = 0$ while keeping $\phi_1$ as a free parameter. 
The complex phases of the trilinear couplings $A_t$, $A_b$, $A_\tau$ and the gaugino-mass parameters $M_i$ 
with $i=1,2,3$ are chosen as default according to
\bea
\phi_{t}=\phi_{b}=\phi_{\tau}=\phi_{3}=\phi_1=\fr{\pi}{2},
\eea
unless specified otherwise.
The phase of $\mu$ is chosen to be zero in order to be consistent with the experimental data  of
the electric dipole moment. We will study the dependence of our results on $\tan\beta$, $M_{H^\pm}$, $\phi_t$ and 
$\phi_3$ in the numerical analysis. The $\phi_b$ dependence is not very interesting since 
it is similar to but much weaker than that of $\phi_t$.

The scale of $\al_s$ in the SUSY-QCD resummation of the effective bottom-Higgs couplings
 \eq{eq:deltamb}  is set to be $Q=(m_{\tilde{b}_1} + m_{\tilde{b}_2} + m_{\tilde{g}})/3$. If not otherwise
specified, we set the renormalization scale equal to the factorization scale, $\mu_R=\mu_F$, in all numerical
results. Our default choice for the factorization scale is $\mu_{F0} = M_W+ M_{H^\pm}$. 

Our study is done for the LHC at $7\tev$ and $14\tev$ center-of-mass energy. In the numerical analysis,
 we will  focus on the latter since the 
total cross section is about an order of magnitude larger. Important results will be shown for both energies.

\subsection{Checks on the results}
\label{sect-ggWH-qcd-gauge}
The results in this paper have been obtained by two independent calculations. 
We have produced, with the help of \fav\ and \fcv\ \cite{Hahn:1998yk}, two different Fortran~77 codes. 
Loop integrals are calculated by using \ltff\ \cite{Hahn:1998yk, ff}. 
The phase-space integration is done by using the Monte Carlo
integrators \bases~\cite{bases} and \vegas~\cite{Lepage:1977sw}. 
The results of the two codes are in full agreement. 
On top, we have also performed a number of other checks: 

For the process $\ggWHpm$, we have verified that the results are QCD gauge invariant. 
This can be easily done in practice by changing the numerical value of 
the gluon polarization vector $\eps_{\mu}(p,q)$, where $p$ is the gluon momentum and $q$ is 
an arbitrary reference vector. QCD gauge invariance means that the squared amplitudes are independent 
of $q$. More details can be found in \cite{Boudjema:2007uh}. 
As already mentioned, we compared our results also to 
the ones of \cite{Brein:2000cv} for the rMSSM and obtained good agreement.

For the process $\bbWHpm$, besides the common checks of UV and IR finiteness, we compared 
our virtual EW corrections to those obtained by using \sloops\ \cite{Baro:2008bg,Baro:2009gn}, 
and the SUSY-QCD corrections 
to the results of Rauch~\cite{Rauch:2008fy} for the case of vanishing phases. 
Again, good agreement was found.

\subsection{\boldmath{$pp/\bbWHpm$}: LO and improved-Born approximations}
\label{sect_results_bbWH_tree}
In this section, we study the effect of the bottom-Higgs coupling resummation described in 
\sect{running_mb} and of the Higgs propagator matrix discussed in 
\sect{sect-ggWH-Higgs-propagator}. 
\begin{figure}[]
\begin{center}
\mbox{\includegraphics[width=0.49\textwidth, height=0.5\textwidth]{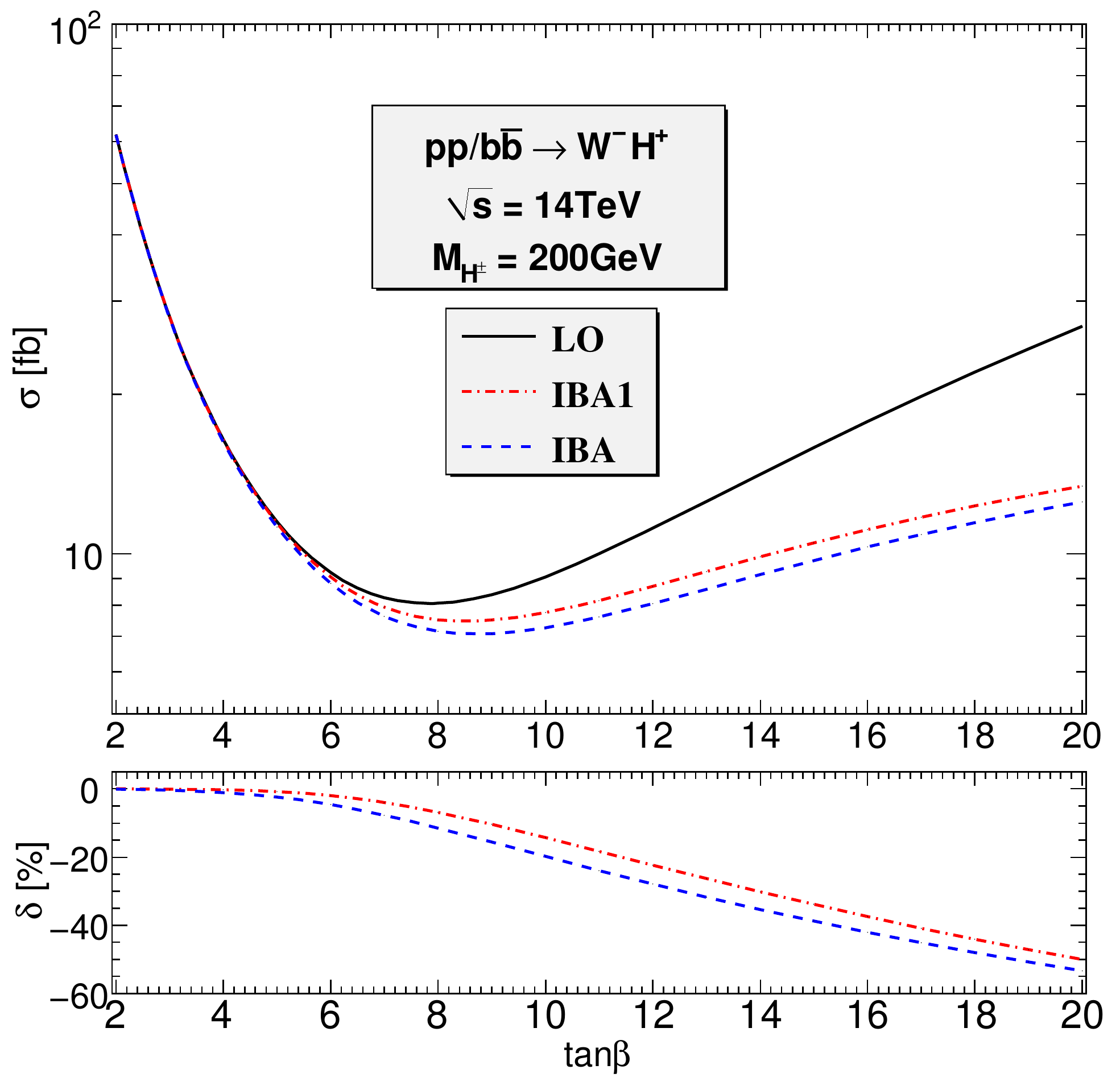}
\hspace*{0.001\textwidth}
\includegraphics[width=0.49\textwidth, height=0.5\textwidth]{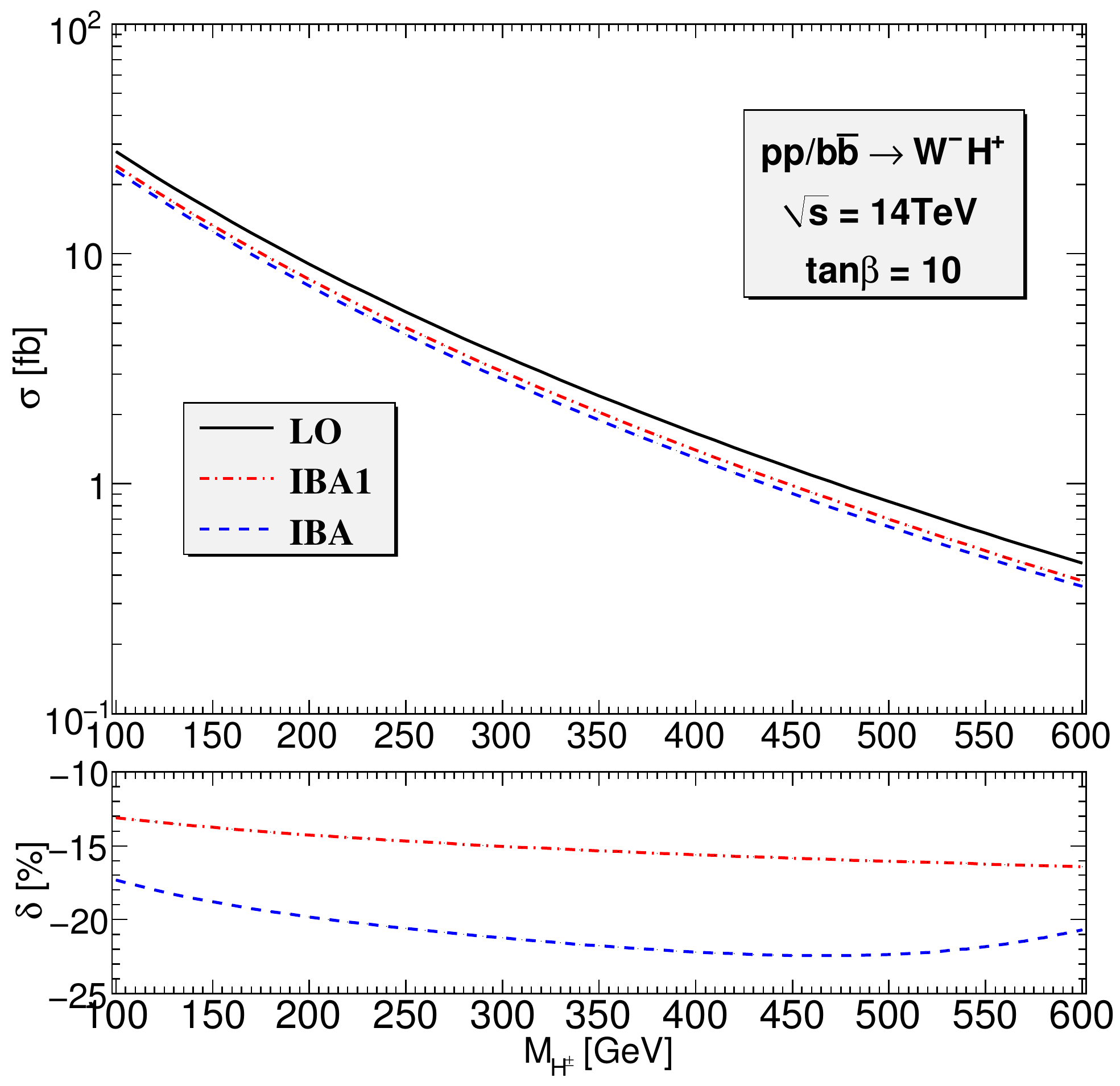}}
\caption{\label{bb_LO_all}{\em The leading order (LO) cross section with $m_b=m_b^{\DRb}$ 
and the two improved Born approximations (IBA) as functions of $\tan\beta$ (left) and $M_{H^\pm}$ (right). 
$\sigma_{IBA1}$ includes the 
$\Delta_b$ resummation but not the Higgs mixing resummation, while $\sigma_{IBA}$ includes both. 
The lower panels show the corresponding relative corrections with respect to the LO result.}}
\end{center}
\end{figure}
\begin{figure}[]
\begin{center}
\mbox{\includegraphics[width=0.49\textwidth, height=0.5\textwidth]{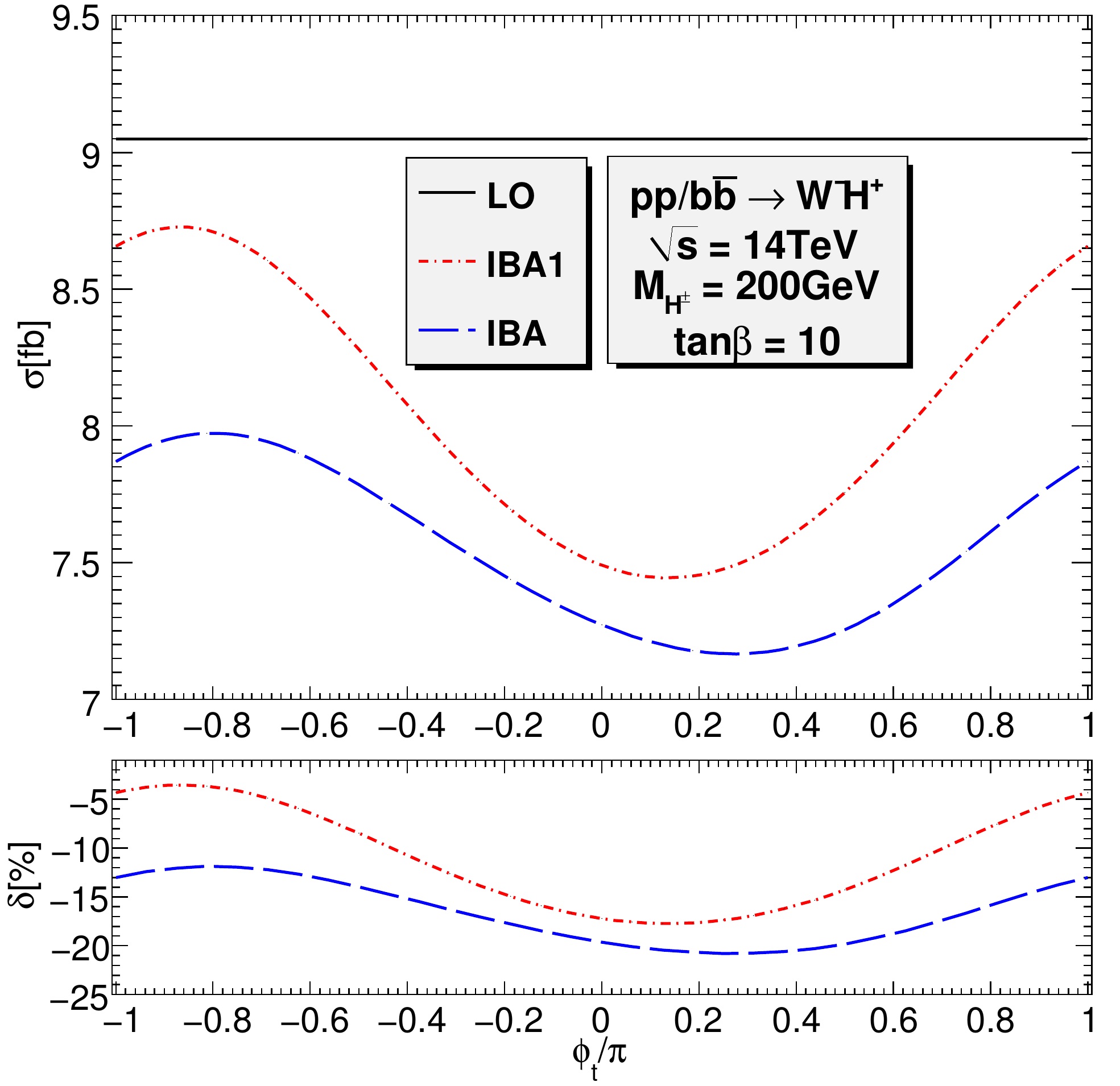}
\hspace*{0.001\textwidth}
\includegraphics[width=0.49\textwidth, height=0.5\textwidth]{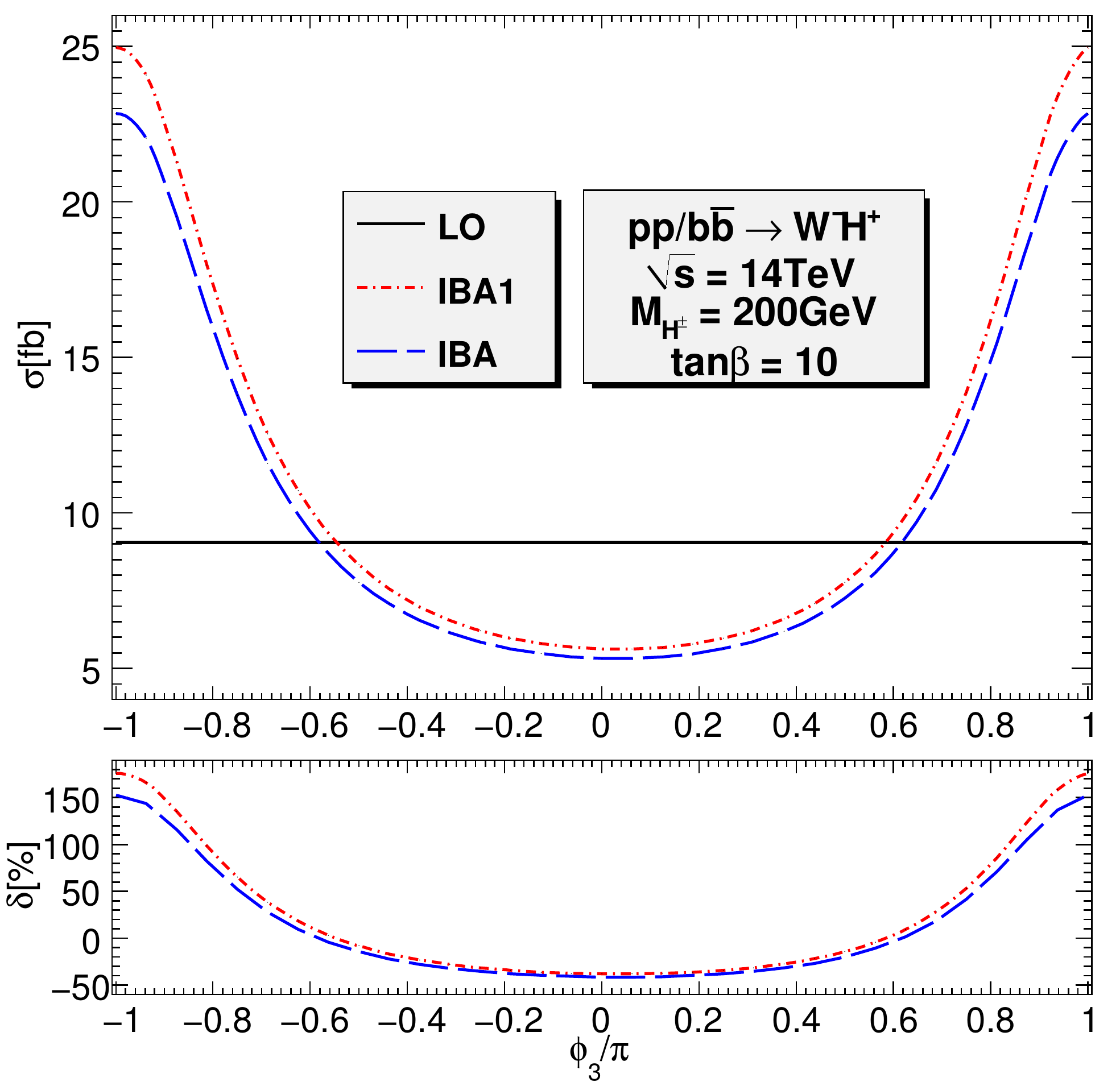}}
\caption{\label{bb_LO_all_phase}{\em Similar to \fig{bb_LO_all}, but with $\phi_t$ (left) and $\phi_3$ (right)
varied instead.}}
\end{center}
\end{figure}

The results for the approximations IBA and IBA1 defined in section \ref{sect-ggWH-Higgs-propagator}
are illustrated in \fig{bb_LO_all} 
showing the dependence on $\tan\beta$ in the left panel and on the mass $M_{H^\pm}$ in the right panel.
The relative correction $\de$, with respect to the LO cross section,
is defined as $\de = (\si_{\text{IBA}} -\si_{\text{LO}})/\si_{\text{LO}}$. 
For small values of $\tan\beta$ the left-chirality  contribution proportional to $m_t/\tan\beta$ is dominant while 
the right-chirality contribution proportional to $m_b\tan\beta$ dominates at large $\tan\beta$. The cross section 
has a minimum around $\tan\beta = 8$.

The effect of $\De_b$ resummation is best understood in terms of \fig{bb_LO_all} and \fig{bb_LO_all_phase}. 
The important point is that $\De_b$ is a complex number and only its real part can interfere with the LO amplitude. 
Thus, the $\De_b$ effect is minimum at $\phi_{t,3}=\pm \pi/2$ where the dominant $\De m_b^{SQCD, \tilde{H}\tilde{t}}$ are purely 
imaginary and is largest at $\phi_{t,3}=0,\pm \pi$. 
$\phi_t$ enters via EW corrections and $\phi_3$ via the SUSY-QCD contributions. 
\fig{bb_LO_all_phase} shows that 
the $\De_b$ effect can be more than $150\%$. In \fig{bb_LO_all} where $\De_b$ is mostly imaginary we see the effect of 
order $\cO(\De_b^2)$ which is about $-15\%$ at $\tan\beta=10$. 
We also observe that the Higgs mixing resummation in 
the $s$-channel diagrams has a much smaller impact, less than $10\%$, as expected.

\subsection{\boldmath{$pp/\bbWHpm$}: full NLO results}
\begin{figure}[]
\begin{center}
\mbox{\includegraphics[width=0.49\textwidth, height=0.5\textwidth]{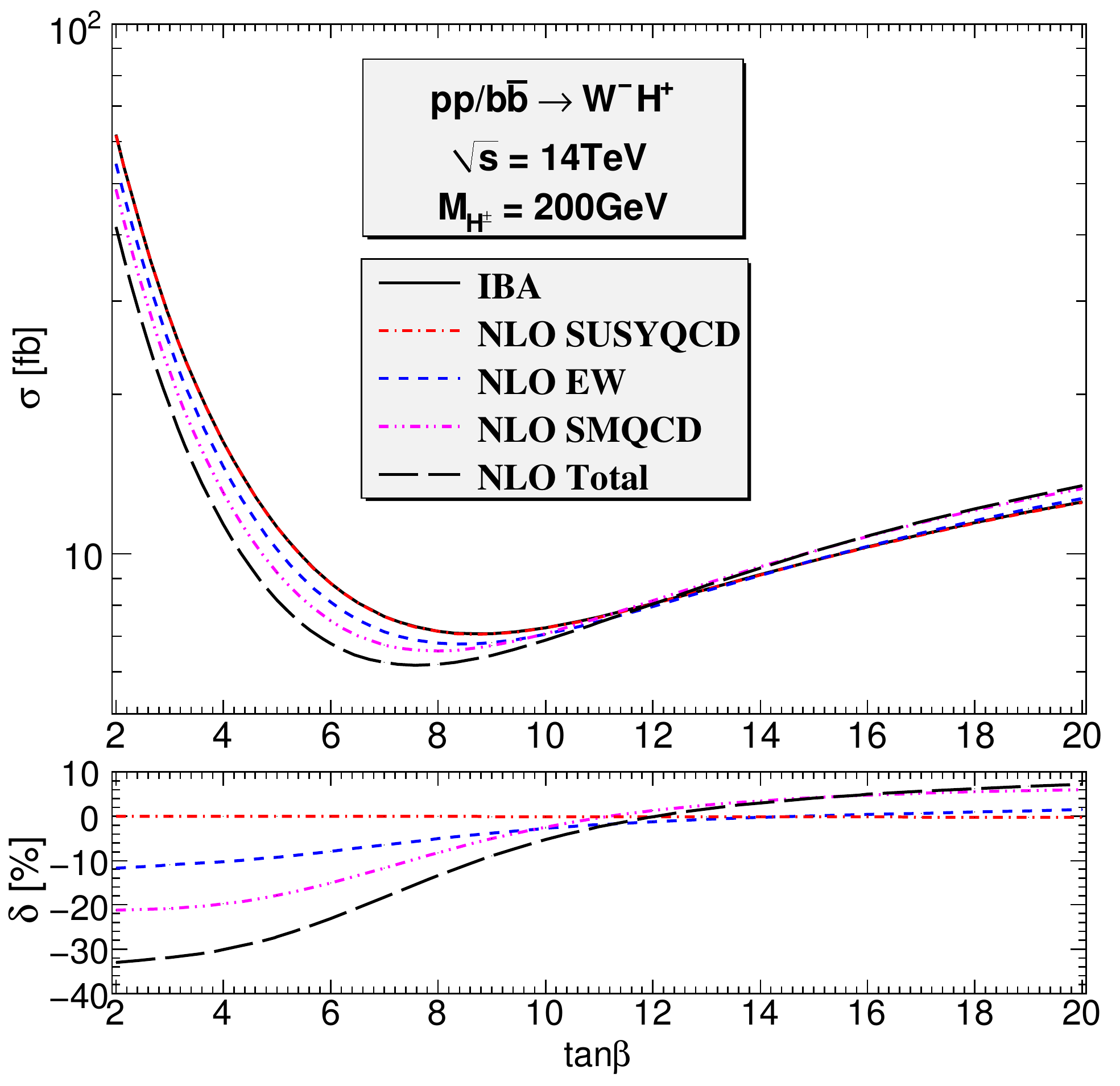}
\hspace*{0.001\textwidth}
\includegraphics[width=0.49\textwidth, height=0.5\textwidth]{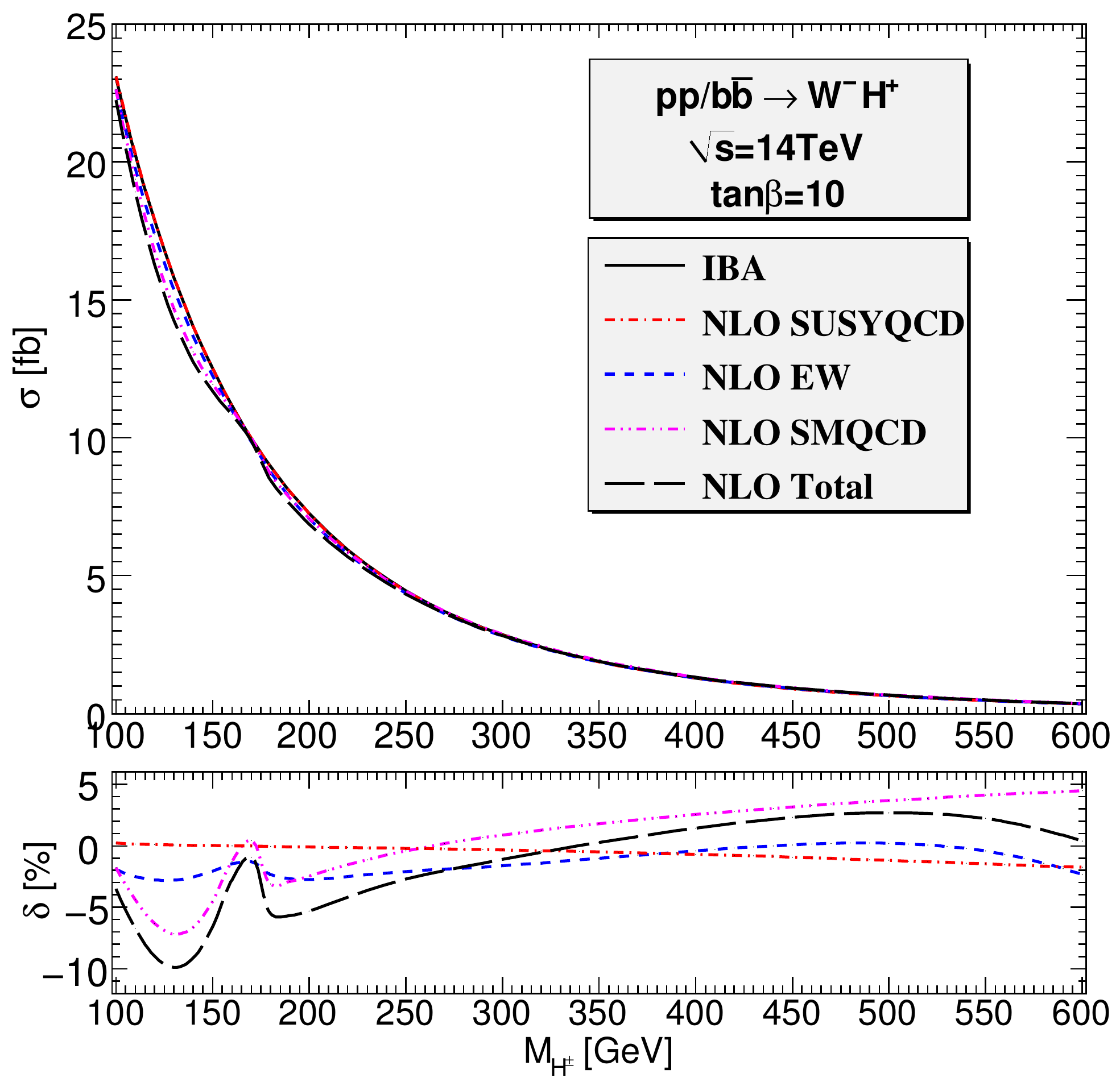}}
\caption{\label{bb_NLO_all}{\em The cross section obtained by using IBA and including various nonuniversal NLO corrections 
as functions of $\tan\beta$ (left) and $M_{H^\pm}$ (right). 
The lower panels show the corresponding relative corrections to the IBA result.}}
\end{center}
\end{figure}
\begin{figure}[]
\begin{center}
\mbox{\includegraphics[width=0.49\textwidth, height=0.5\textwidth]{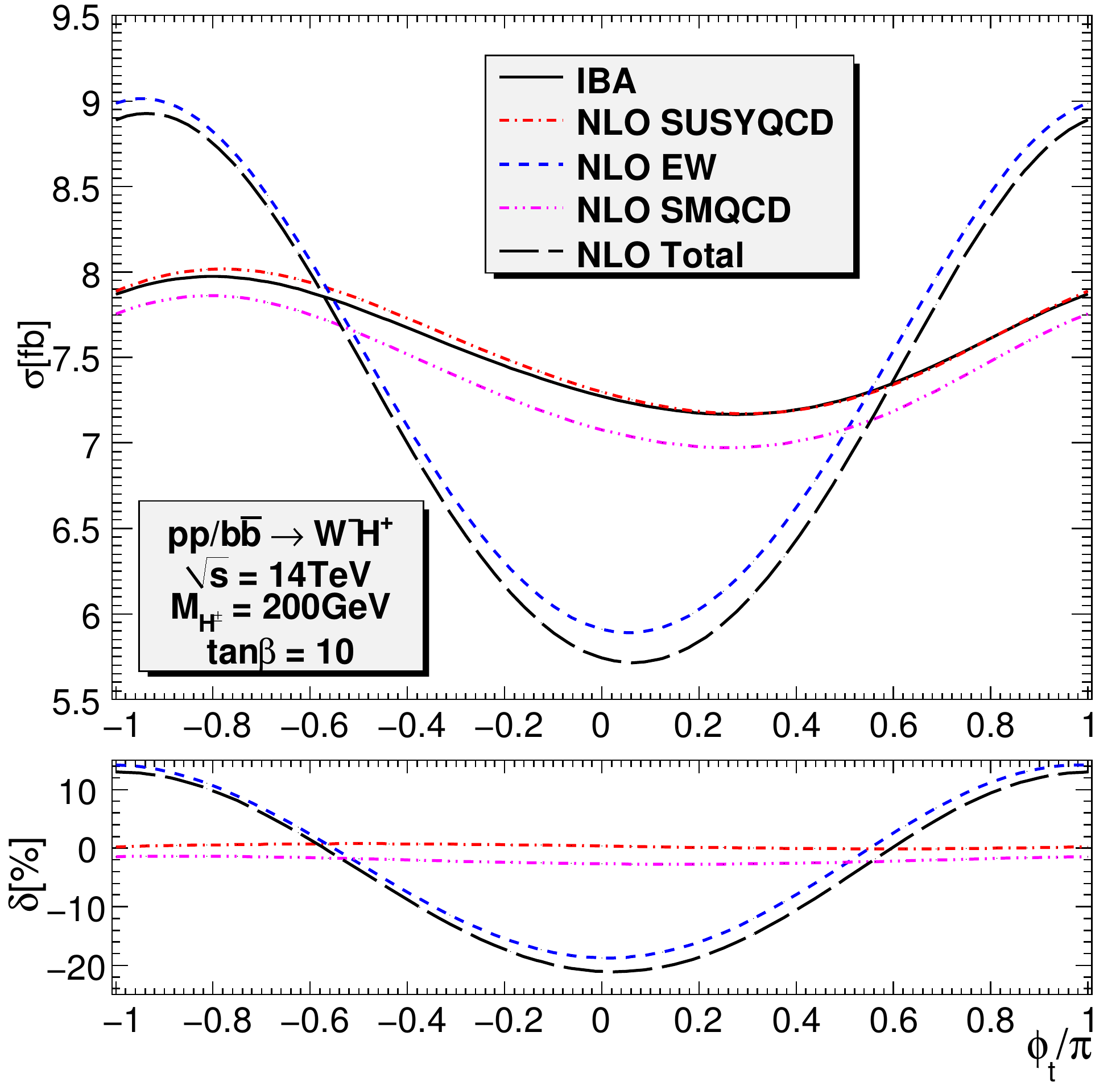}
\hspace*{0.001\textwidth}
\includegraphics[width=0.49\textwidth, height=0.5\textwidth]{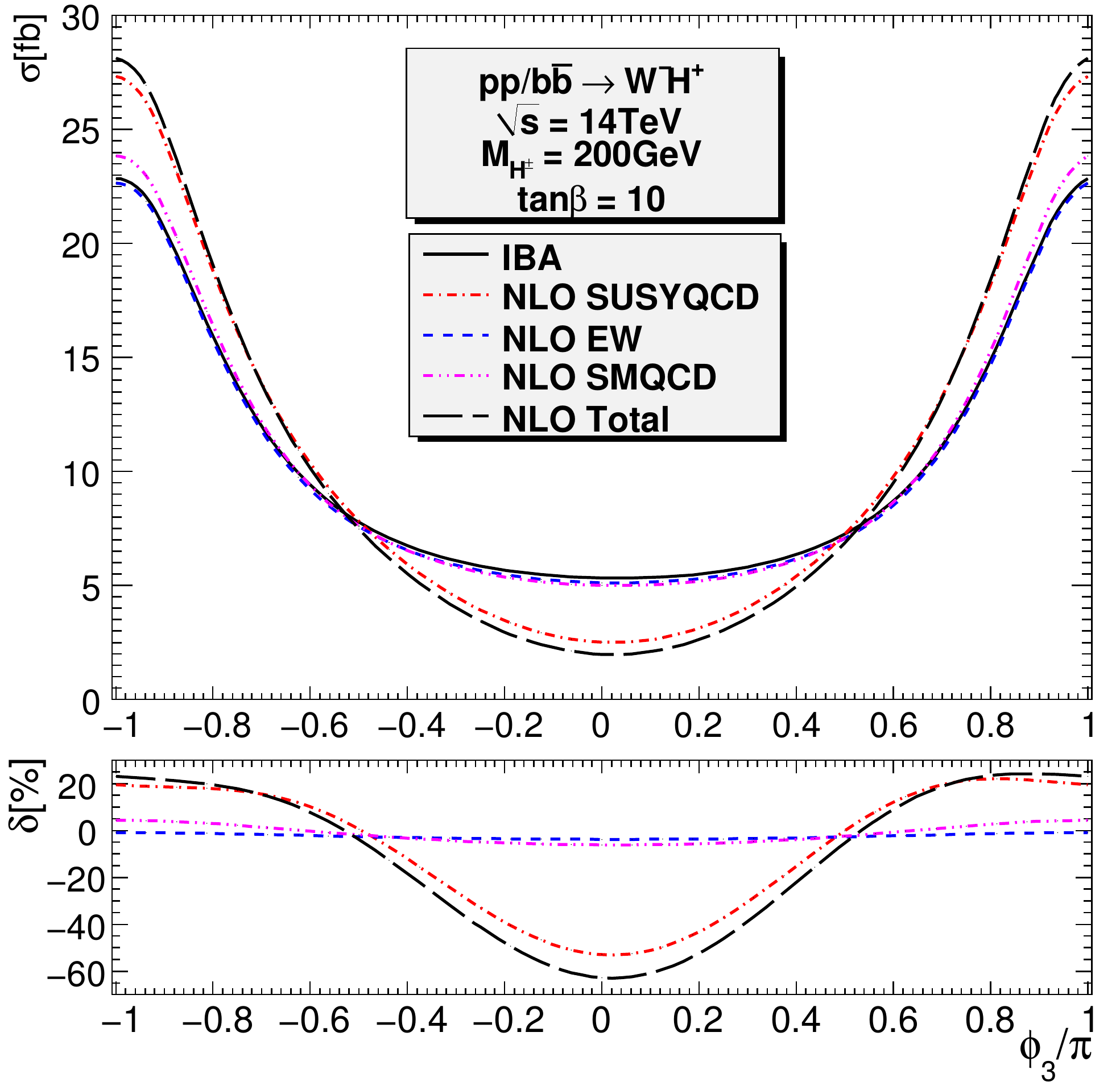}}
\caption{\label{bb_NLO_all_phase}{\em Similar to \fig{bb_NLO_all}, but with $\phi_t$ (left) and $\phi_3$ (right) 
varied instead.}}
\end{center}
\end{figure}
In this section, we investigate the effects of the SUSY-QCD, SM-QCD, and EW  contributions at NLO. 
As in the previous section, we present here two sets of plots. In 
\fig{bb_NLO_all} we show the dependence of the total cross sections on $\tan\beta$ and $M_{H^\pm}$ 
at the default CPX phases, in particular $\phi_{t}=\phi_{3}=\pi/2$. As explained above, 
the $\cO(\De_b)$ effect is turned off in this CPX scenario. The SUSY-QCD and EW  NLO terms
 are therefore small at large $\tan\beta $, as shown in \fig{bb_NLO_all} (left). 
 The SM-QCD correction is about $-20\%$ for small $\tan\beta$ and changes 
 the sign around $\tan\beta =11$ due to the competition between 
 the $b\bar{b}$ and the $g$-induced contributions. All the NLO contributions for
 different values of $\tan\beta$ and $M_{H^\pm}$ 
can be found in \tab{table_NLO}.
\fig{bb_NLO_all_phase} shows the dependence of the total cross sections 
on $\phi_t$ and $\phi_3$ for $\tan\beta=10$ and $M_{H^\pm}=200$GeV. 
The EW corrections depend strongly on $\phi_t$, and the SUSY-QCD corrections 
on $\phi_3$. At $\phi_{t}=\phi_{3}=0,\pm\pi$ the effects are largest. 
The remaining EW and SUSY-QCD corrections,  
beyond the $\cO(\De_b)$ contribution,
are still rather large.
\begin{figure}[t]
  \centering
  \includegraphics[width=0.6\textwidth]{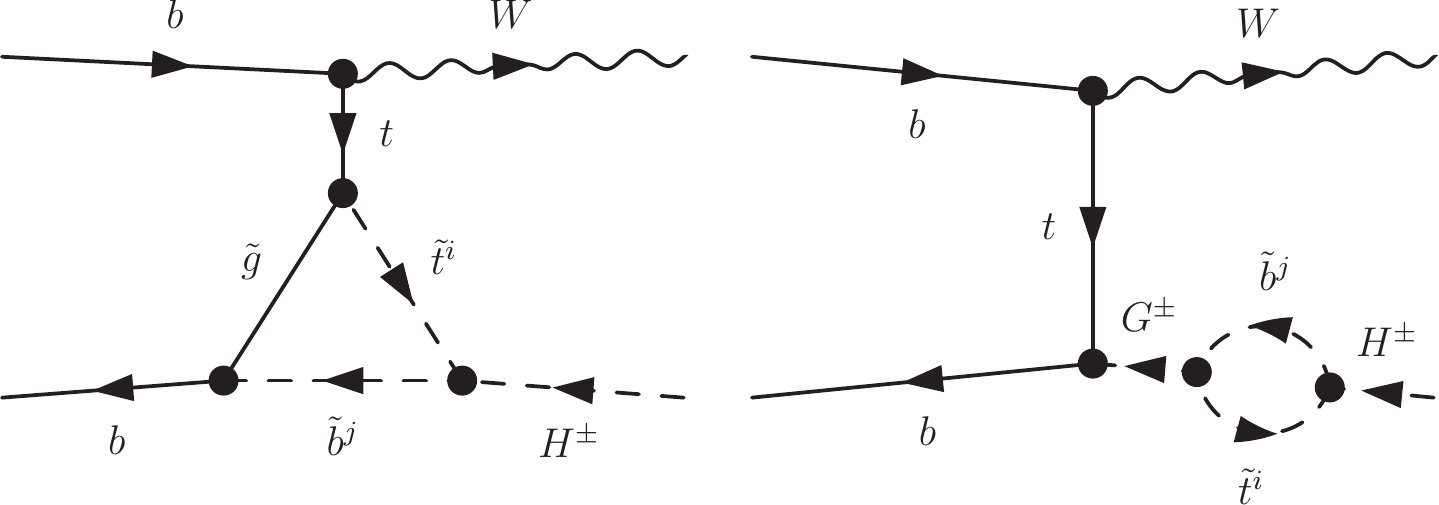}
  \caption{{\em Diagrams that can introduce large SUSY-QCD (left) and EW (right) corrections. $G^\pm$ are the charged Goldstone bosons.}}
  \label{diag_tbH_subleading}
\end{figure}
In particular, there is the following term of the SUSY-QCD correction,  
\bea
\tilde\De_{t}&=&\fr{2\alpha_s}{3\pi}M_3^*\mu^*\tan\beta J(m_{\tilde{g}}^2),\crn
J(m^2)&=&\vert U_{11}^{\tilde b}\vert^2 \vert U_{12}^{\tilde t}\vert^2 I(m^2, m_{\tilde{t}_1}^2, m_{\tilde{b}_1}^2) 
+ \vert U_{21}^{\tilde b}\vert^2 \vert U_{12}^{\tilde t}\vert^2 I(m^2, m_{\tilde{t}_1}^2, m_{\tilde{b}_2}^2)\crn 
&+& \vert U_{11}^{\tilde b}\vert^2 \vert U_{22}^{\tilde t}\vert^2 I(m^2, m_{\tilde{t}_2}^2, m_{\tilde{b}_1}^2)
+ \vert U_{21}^{\tilde b}\vert^2 \vert U_{22}^{\tilde t}\vert^2 I(m^2, m_{\tilde{t}_2}^2, m_{\tilde{b}_2}^2), 
\label{eq:sub_leading_Htb}
\eea
which can be included in the top-Yukawa part of charged Higgs couplings as follows
\bea
\tilde\lambda_{b\bar{t}H^+}&=&\fr{ie}{\sqrt{2}s_WM_W}\left(\fr{m_t}{\tbeta}(1-\tilde\De_{t})P_L 
+ \mbDRb\tbeta \Delta_b^{3*}P_R\right),\crn
\tilde\lambda_{t\bar{b}H^-}&=&\fr{ie}{\sqrt{2}s_WM_W}\left(\mbDRb\tbeta \Delta_b^{3}P_L 
+ \fr{m_t}{\tbeta}(1-\tilde\De^{*}_{t}) P_R\right).
\label{eq:sub_leading_Htb_couplings}
\eea
This term originates from the left diagram in \fig{diag_tbH_subleading} and is important for small $\tan\beta$.
This finding agrees with the discussion in \cite{Carena:2002bb} where other subleading corrections are also discussed. 
If the couplings \eq{eq:sub_leading_Htb_couplings} are used we find that 
the new-improved LO results move significantly closer to the full NLO results in \fig{bb_NLO_all_phase} (right). 
The situation in the left part of \fig{bb_NLO_all_phase} is due to the EW corrections. 
It indicates that there are 
still large corrections proportional to $A_t\mu\alpha_t/(4\pi)$ 
which can be associated with the right diagram in \fig{diag_tbH_subleading}.

The SM-QCD corrections (and EW corrections to a lesser extend) have a striking
structure for small masses $M_{H^\pm} < m_t$ (\fig{bb_NLO_all}, right part). 
This is due to the finite contribution of the process $\bgluWHpb$. When $M_{H^\pm} < m_t$ the 
intermediate top quark can be on-shell and can decay to $H^+b$. As discussed in \sect{sect-subtraction-OS}, this OS contribution 
has to be properly subtracted. The structure indicates that the OS top-quark effect cannot be completely removed 
and this quantum effect on the $W^-H^+$ production rate is an interesting feature, which was not discussed in previous studies 
\cite{Hollik:2001hy, Gao:2007wz}.
\subsection{\boldmath{$pp/\ggWHpm$}: neutral Higgs-propagator effects}
\label{sect_results_ggWH}
\begin{figure}[]
\begin{center}
\mbox{\includegraphics[width=0.49\textwidth, height=0.5\textwidth]{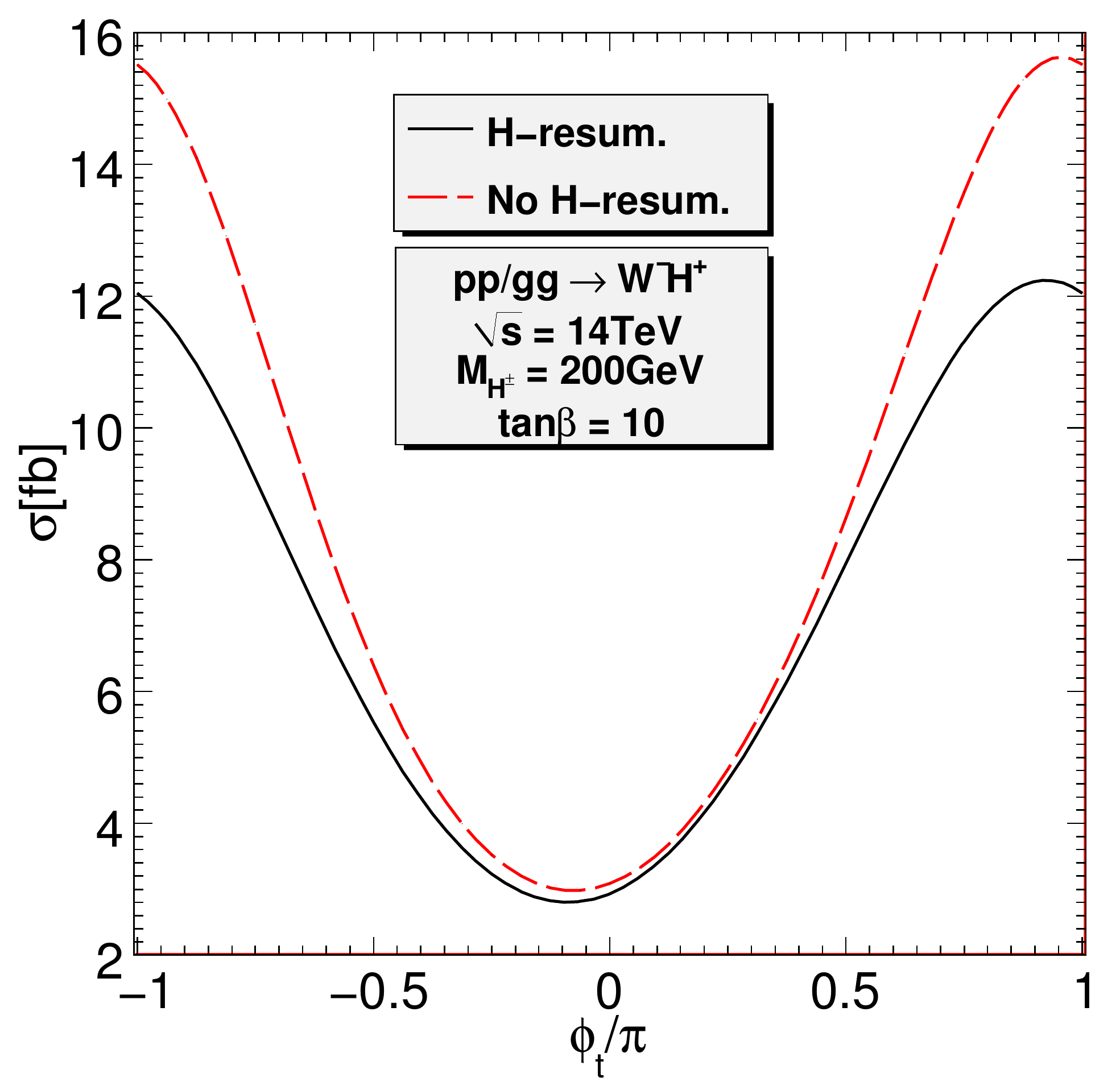}
\hspace*{0.001\textwidth}
\includegraphics[width=0.49\textwidth, height=0.5\textwidth]{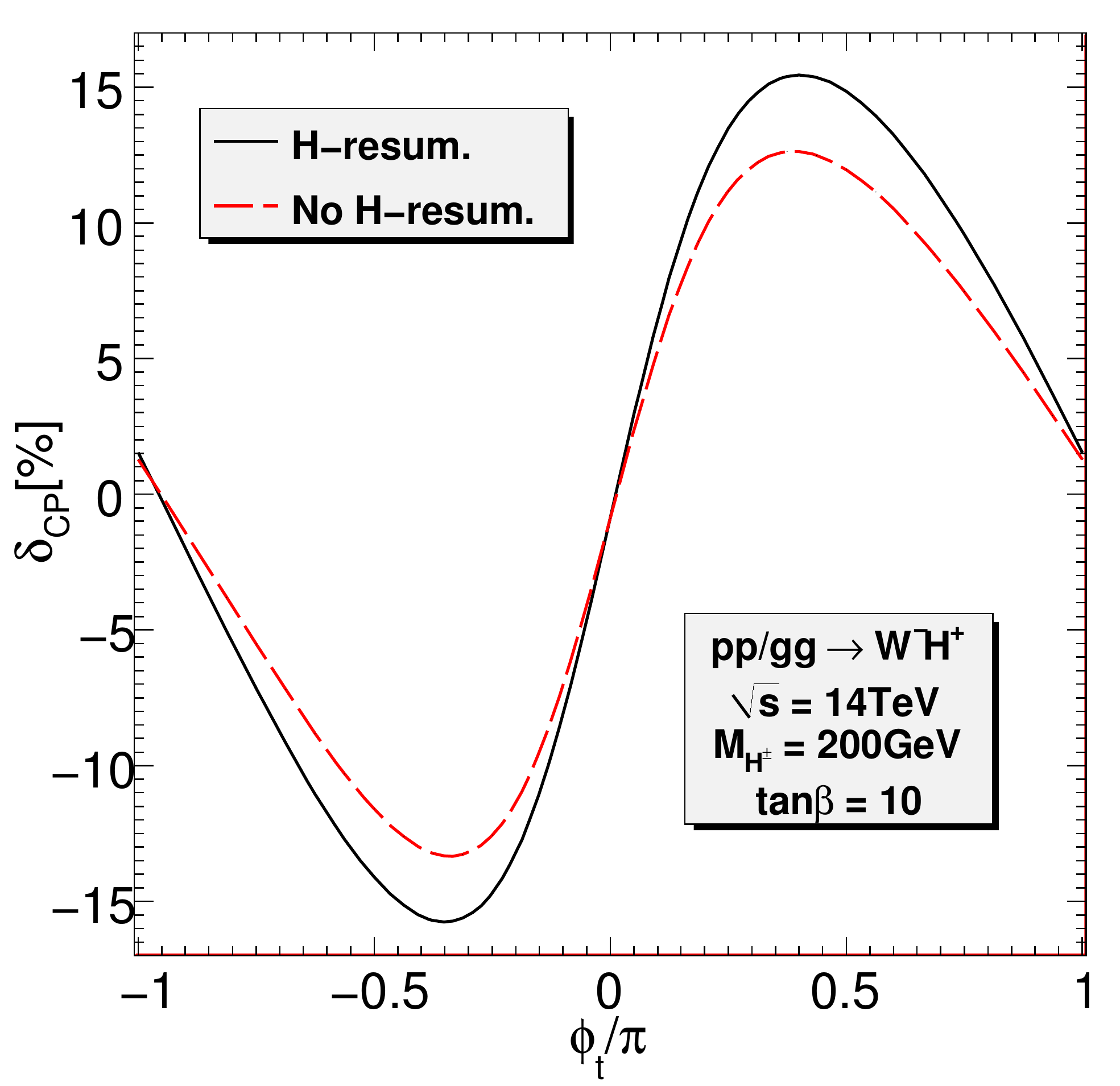}}
\caption{\label{gg_PHITP}{\em The cross section (left) and CP asymmetry (right) as functions of $\phi_{t}$.}}
\end{center}
\end{figure}
Even though the $gg$-fusion subprocess is loop induced, its contribution can be  
of the same order as the tree-level $\bbWHpm$ contribution. 
Neutral Higgs bosons are exchanged in the $s$-channel and can be described by using
effective bottom-Higgs couplings and the full Higgs-propagator matrix.
The impact of the latter on the total cross section and CP asymmetry is 
large as can be seen from \fig{gg_PHITP}. 
The cross section can be reduced by  $20\%$ at $\phi_t=\pm\pi$, while the CP asymmetry increases about 
$25\%$ at $\phi_t=\pm\pi/2$. This is consistent with the discussion in \sect{sect-ggWH-Higgs-propagator}. 
We also observe that the $gg$ contribution is very sensitive to $\phi_t$. 
\subsection{\boldmath{$\ppWHpm$}: total results at $7\tev$ and $14\tev$}
\begin{figure}[]
\begin{center}
\mbox{\includegraphics[width=0.49\textwidth, height=0.5\textwidth]{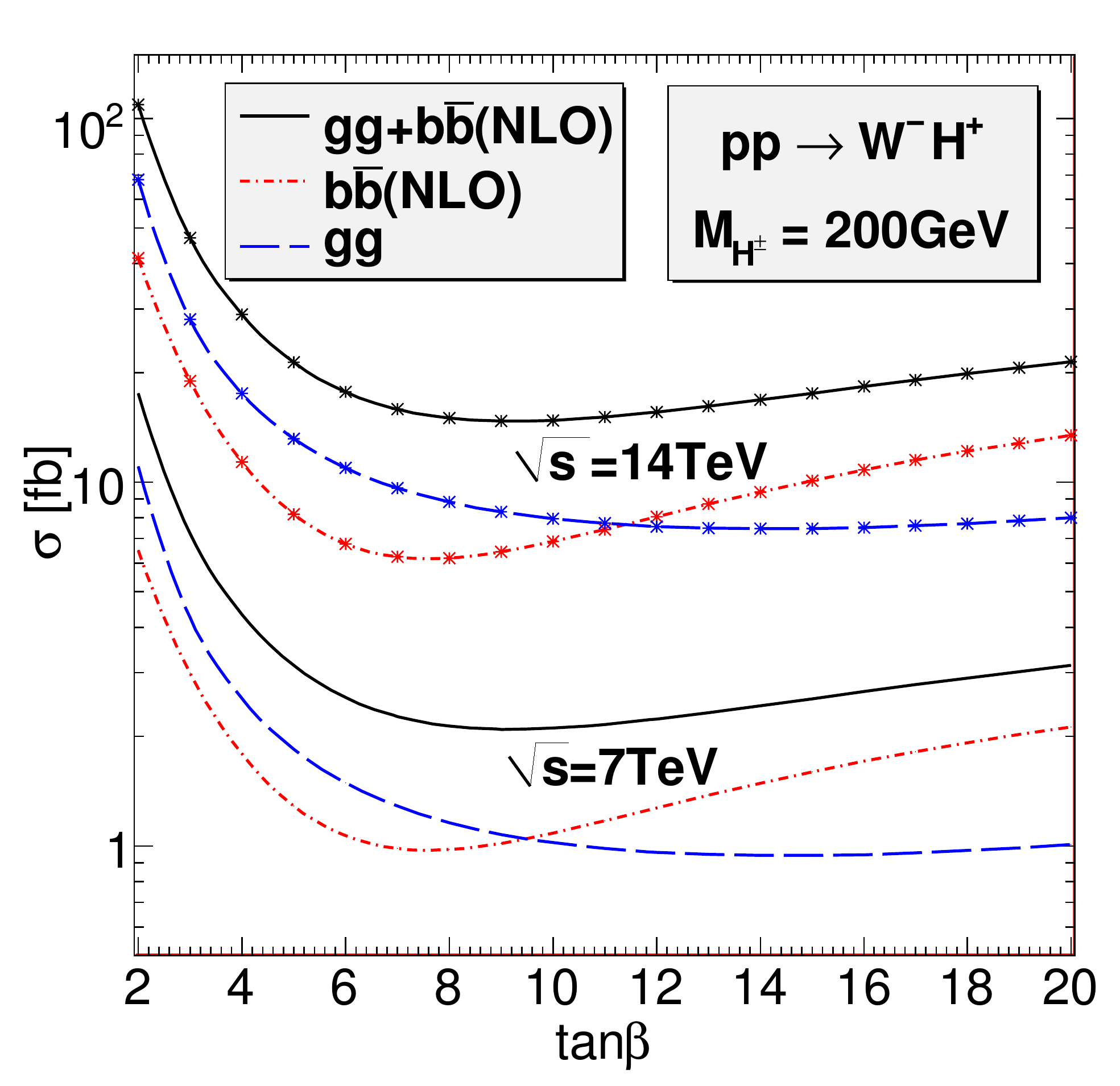}
\hspace*{0.001\textwidth}
\includegraphics[width=0.49\textwidth, height=0.5\textwidth]{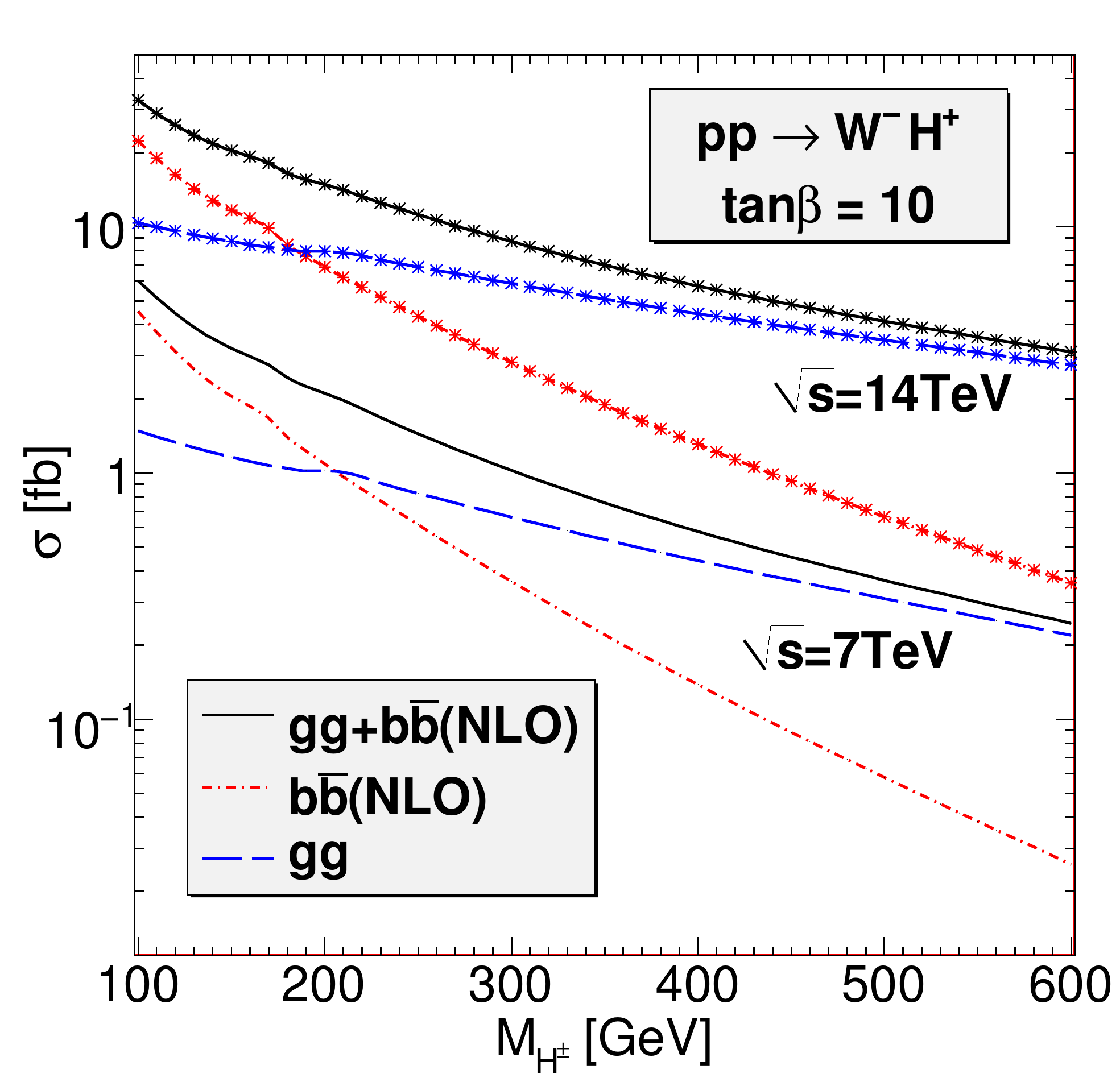}}
\caption{\label{pp_NLO_all}{\em The cross section as a function of $\tan\beta$ (left) and $M_{H^\pm}$ (right).}}
\end{center}
\end{figure}

\begin{figure}[]
\begin{center}
\mbox{\includegraphics[width=0.49\textwidth, height=0.5\textwidth]{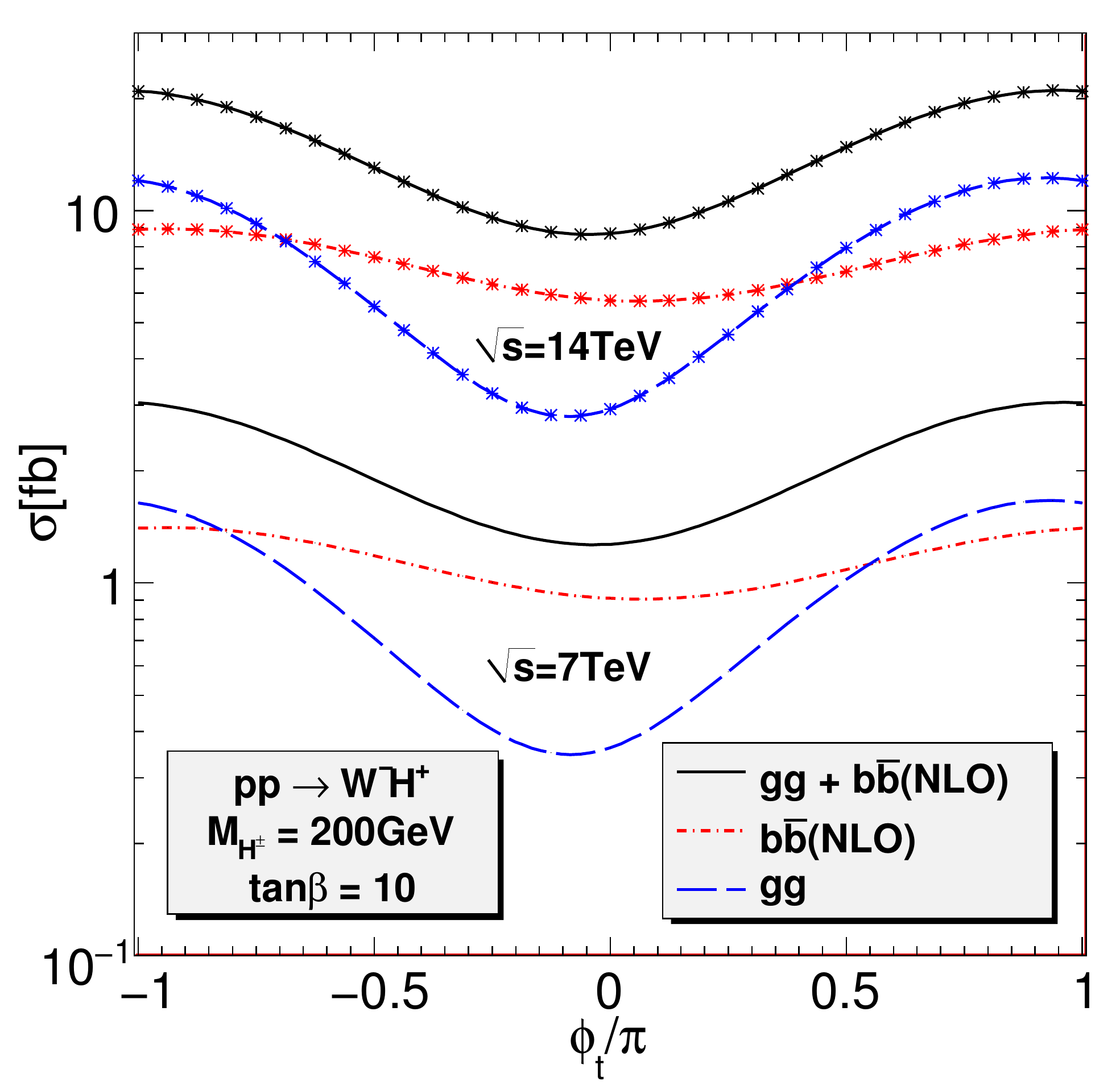}
\hspace*{0.001\textwidth}
\includegraphics[width=0.49\textwidth, height=0.5\textwidth]{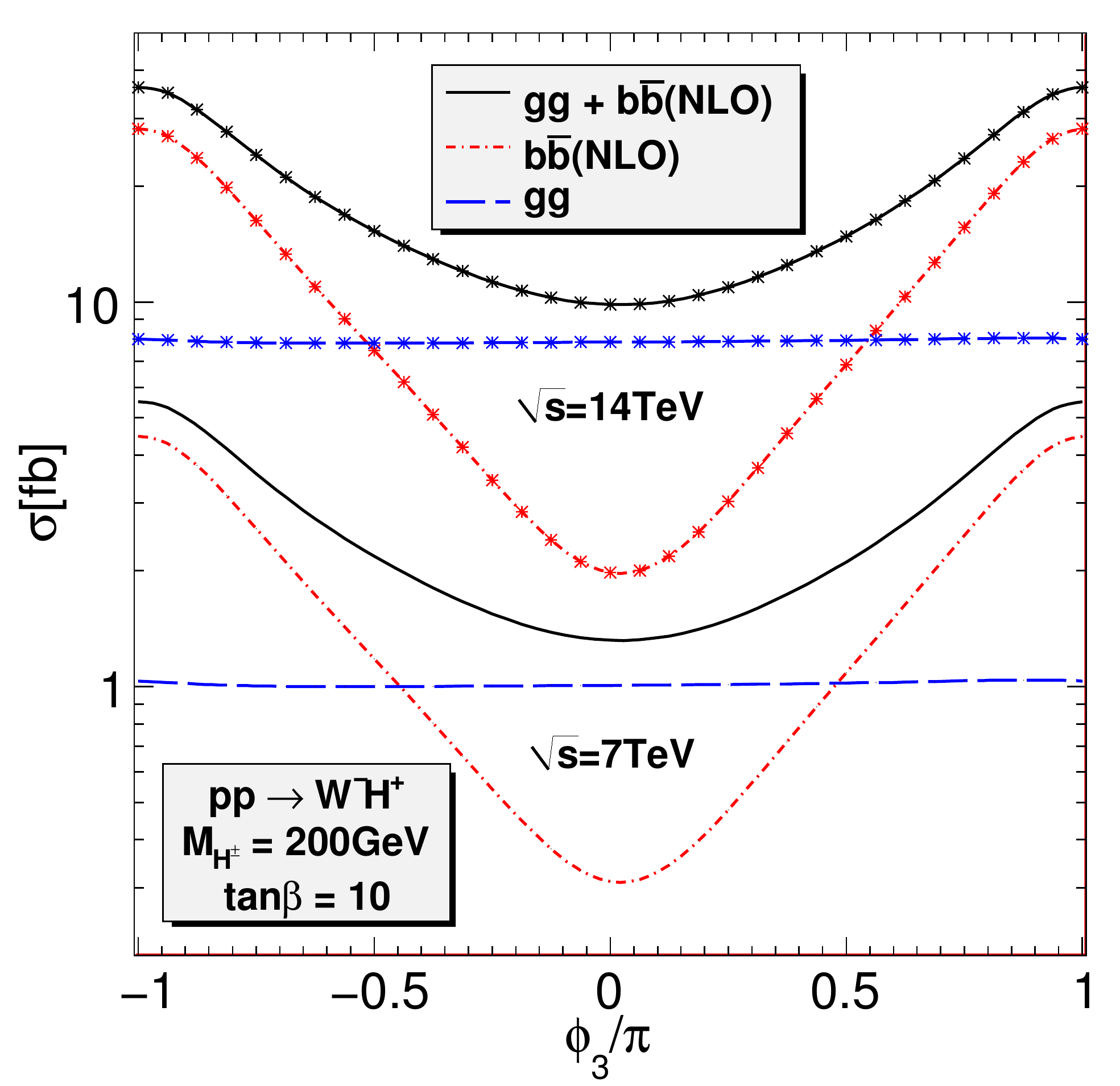}}
\caption{\label{pp_NLO_all_phase}{\em The cross section as a function of $\phi_{t}$ (left) and $\phi_{3}$ (right).}}
\end{center}
\end{figure}

The total production cross section for the $W^-H^+$ final state at the LHC is shown in \fig{pp_NLO_all} and 
\fig{pp_NLO_all_phase}, as well as in  \tab{table_NLO}. The 
cross section increases by an order of magnitude when the center-of-mass energy goes from $7\tev$ to 
$14\tev$. The $gg$ contribution is largest for small $\tan\beta$ and large $M_{H^\pm}$ 
while the $b\bar{b}$ dominates when $\tan\beta > 12$ and, approximately, $M_{H^\pm}<200\gev$. In the right panel
of \fig{pp_NLO_all}, one can see a little bump on the $gg$ contribution  around $M_{H^\pm}=200$GeV,
attributed  to the three-point Landau singularities discussed in \sect{sect-ggWH}.
The total cross section depends strongly on the phases $\phi_t$ and $\phi_3$ as can be seen from 
\fig{pp_NLO_all_phase}. The $gg$ contribution is almost independent of $\phi_3$ since the gluino does 
not appear at the one-loop level (the contribution through $\De_b$ resummation is of higher-order effect).

\begin{figure}[]
\begin{center}
\mbox{\includegraphics[width=0.49\textwidth, height=0.5\textwidth]{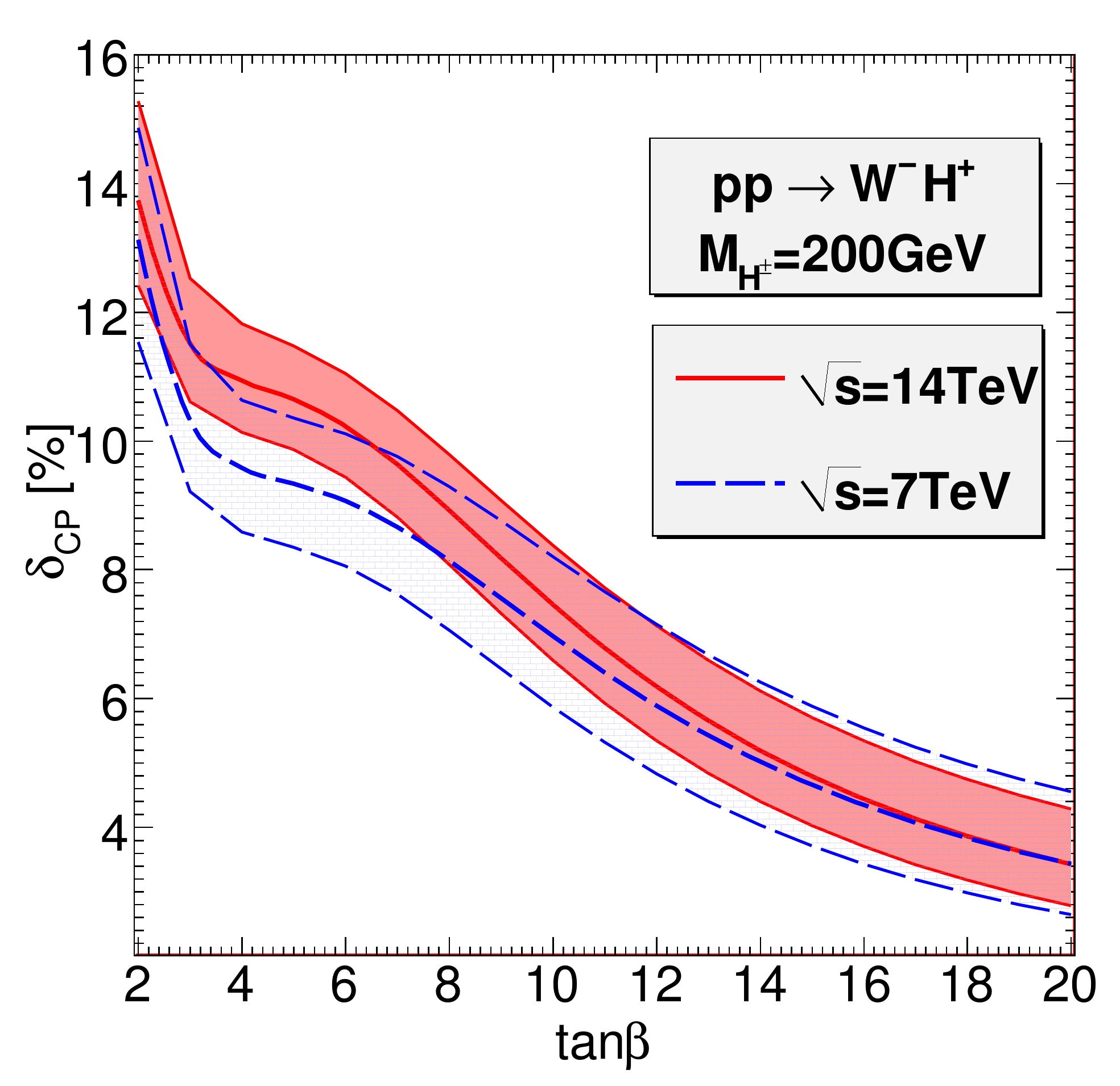}
\hspace*{0.001\textwidth}
\includegraphics[width=0.49\textwidth, height=0.5\textwidth]{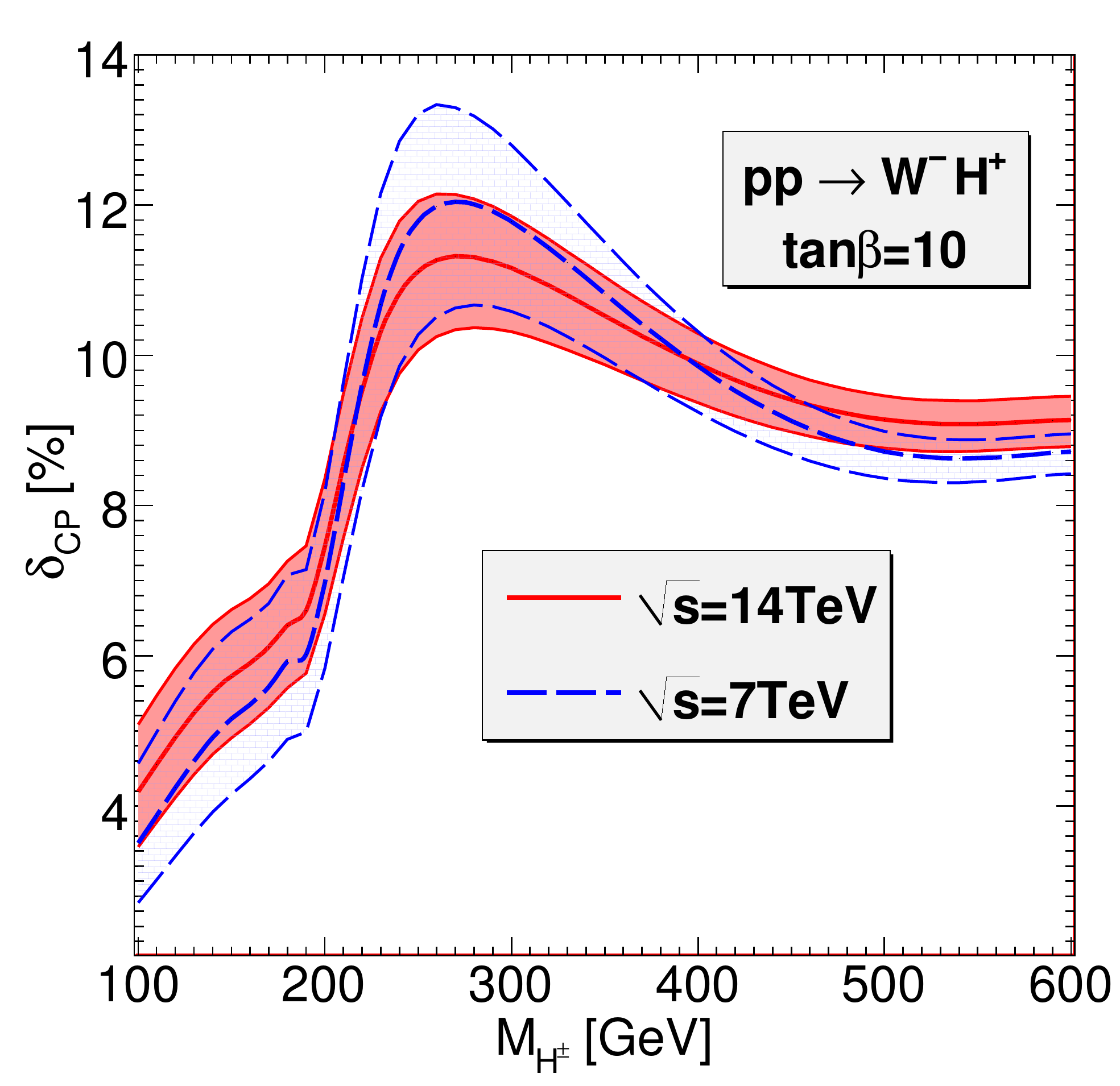}}
\caption{\label{pp_NLO_CP}{\em CP asymmetry as a function of $\tan\beta$ (left) and $M_{H^\pm}$ (right). Within the band,
the scale $\mu_R=\mu_F$ is varied in the range $\mu_{F0}/2<\mu_F<2\mu_{F0}$.}}
\end{center}
\end{figure}

\begin{figure}[]
\begin{center}
\mbox{\includegraphics[width=0.49\textwidth, height=0.5\textwidth]{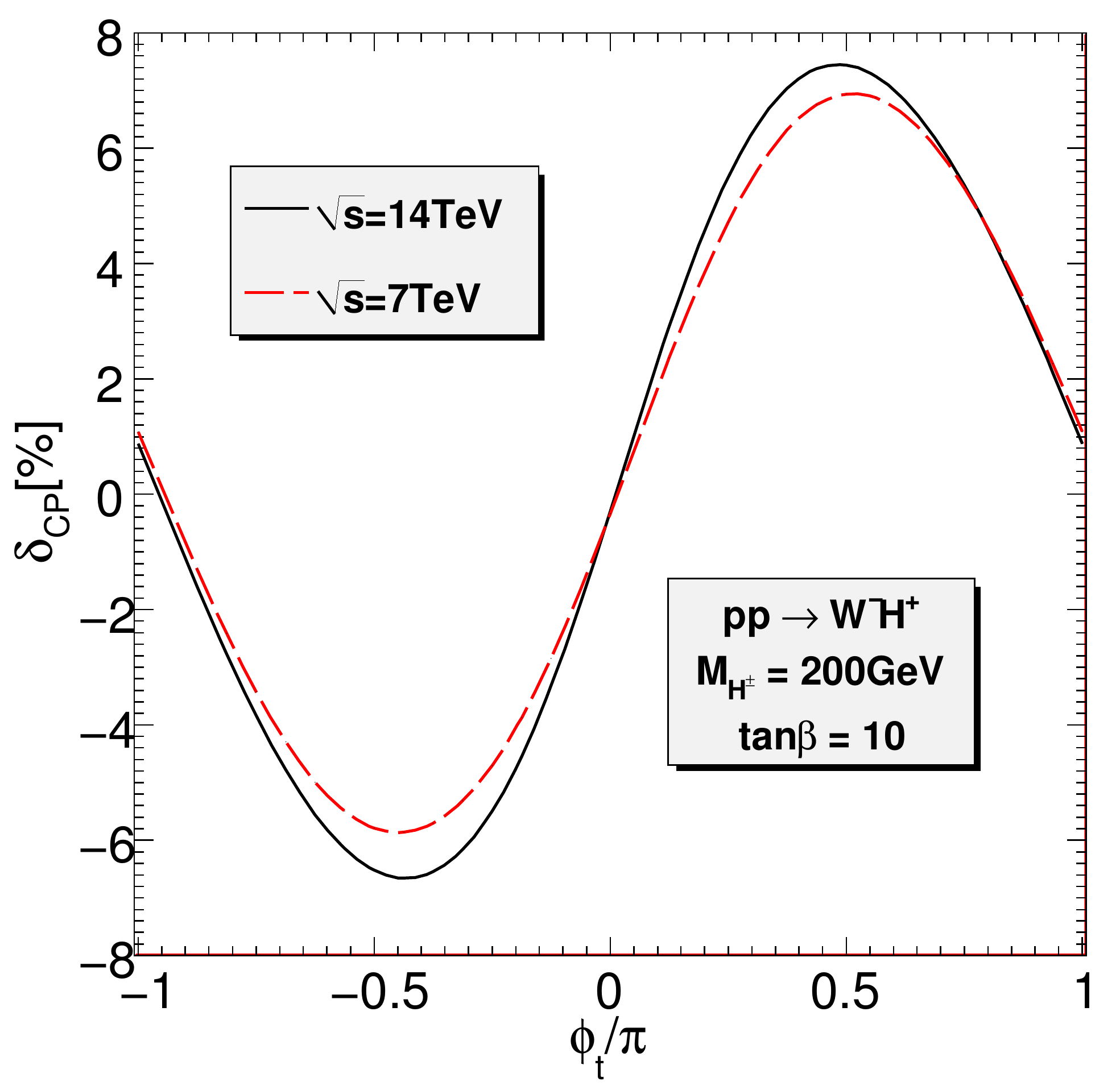}
\hspace*{0.001\textwidth}
\includegraphics[width=0.49\textwidth, height=0.5\textwidth]{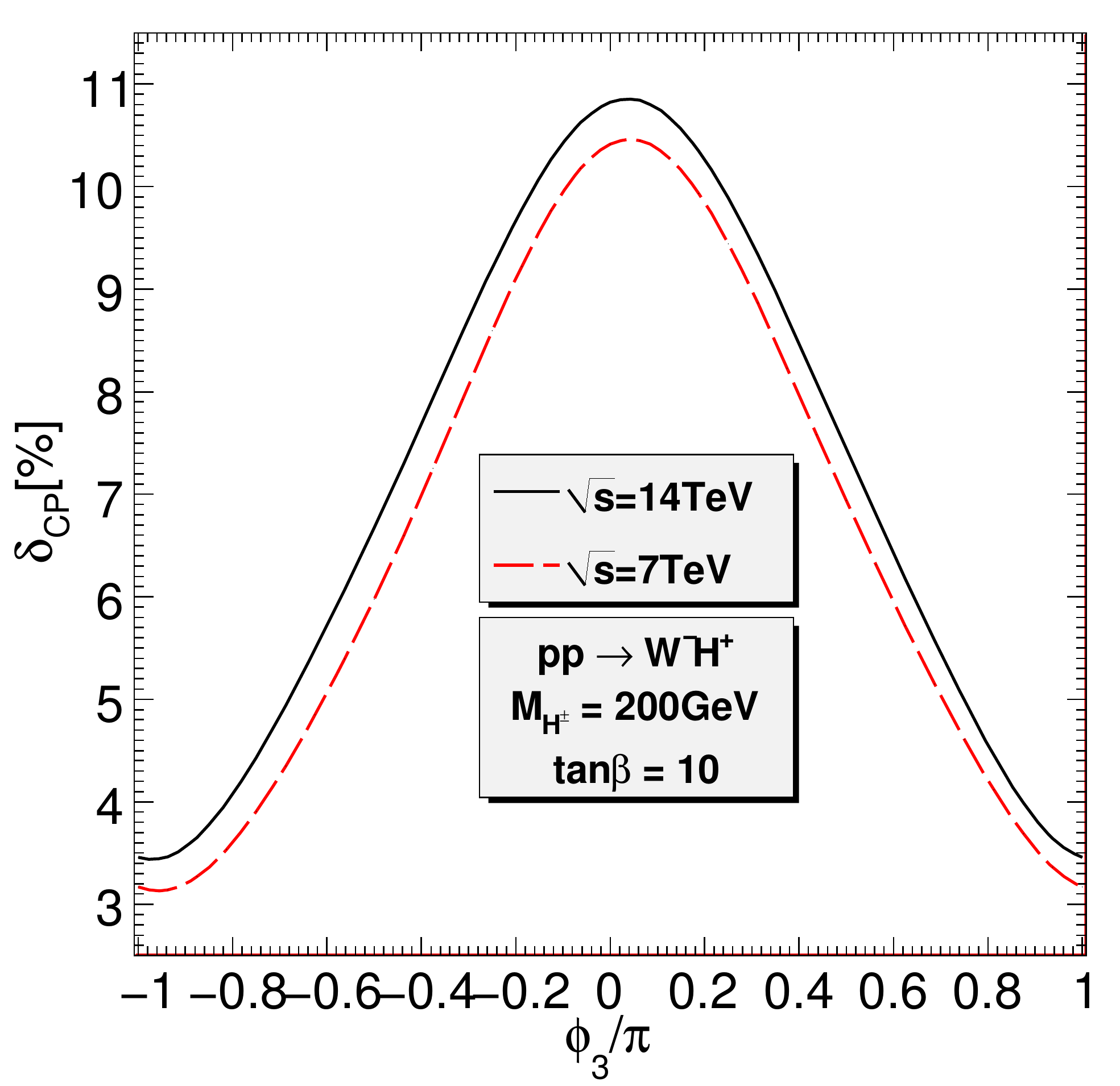}}
\caption{\label{pp_NLO_CP_phase}{\em CP asymmetry as a function of $\phi_{t}$ (left) and $\phi_{3}$ (right).}}
\end{center}
\end{figure}

The CP violating asymmetry is shown in \fig{pp_NLO_CP} as a function of $\tan\beta$ and $M_{H^\pm}$, 
and in \fig{pp_NLO_CP_phase} versus $\phi_t$ and $\phi_3$. 
The uncertainty bands obtained by varying the renormalization 
and factorization scales (we set $\mu_R=\mu_F$ for simplicity) in the 
range $\mu_{F0}/2 < \mu_F < 2\mu_{F0}$ are shown only in \fig{pp_NLO_CP} 
since the uncertainty depends strongly on $\tan\beta$ and in particular on
$M_{H^\pm}$, but not on the phases. A more detailed account of the scale uncertainty 
of our results is given in the next section.  
As discussed at the end of \sect{sec:hadronic}, the CP violating effect is dominantly generated 
by the gluon-gluon fusion channel. 
The $b\bar{b}$ channel contributes significantly to the symmetric cross
section and thus to the denominator of 
the CP asymmetry. It is therefore easy to understand why $\delta_{CP}$ is small for large $\tan\beta$ and 
small $M_{H^\pm}$, as seen in \fig{pp_NLO_CP}. The dependence on $\phi_3$ is explained by the same reasons: the numerator is 
independent of $\phi_3$ while the denominator including $\sigma_{b\bar{b}}$ has a 
minimum at $\phi_3=0$. The CP asymmetry is therefore maximum around $\phi_3=0$.

\begin{table}[]
 \begin{footnotesize}
 \bc 
 \caption{\label{table_NLO}{
\em The total cross section in$\fb$ for $pp/bb \to W^- H^+$ including the IBA
and various nonuniversal NLO corrections and 
for $pp/gg \to  W^- H^+ $ at $\sqrt{s}= 14 \tev$.
The charged Higgs-boson masses are given in$\gev$.}}
\vspace*{0.5cm}
\begin{tabular}{l c r@{.}l r@{.}l r@{.}l r@{.}l r@{.}l r@{.}l}
 \hline
$\tan\beta$
& $M_{H^\pm}$ 
&\multicolumn{2}{c}{ $\si_{\text{IBA}}$}
&\multicolumn{2}{c}{ $\De_{\text{EW}}$}
&\multicolumn{2}{c}{$\De_{\text{SMQCD}}$}
&\multicolumn{2}{c}{$\De_{\text{SUSYQCD}}$}
&\multicolumn{2}{c}{ $\si_{gg}$}
&\multicolumn{2}{c}{all}\\
\hline \hline
5  & 200 & 11&241(1)  & -1&0383(3)  &  -2&012(3)  & -0&00821(1) & 13&194(1) &  21&377(3) \\ 
10 & 200 & ~7&2568(9) & -0&1989(5)  &  -0&178(1)  & -0&00721(2) & ~7&9428(5)&  14&815(2) \\
20 & 200 & 12&546(2)  & 0&1881(6)  &  ~0&752(3)  & -0&03570(6) & ~7&9968(6)&  21&447(4) \\
10 & 150 & 12&497(1)  & -0&2574(5)  &  -0&561(2)  & 0&00191(4)  & ~8&7064(5)&  20&387(3) \\
10 & 400 & ~1&2907(2) & -0&00530(7) &  ~0&0328(2) & -0&008954(7)& ~4&4386(3)&  ~5&7477(4) \\
10 & 600 & ~0&35740(5)& -0&00832(2) &  ~0&01594(5)& -0&006263(4)& ~2&7481(1)&  ~3&1069(2) \\ 
\hline 
\end{tabular}\ec
 \end{footnotesize}
\end{table} 


\subsection{Scale dependence}
\begin{figure}[]
\begin{center}
\mbox{\includegraphics[width=0.49\textwidth, height=0.5\textwidth]{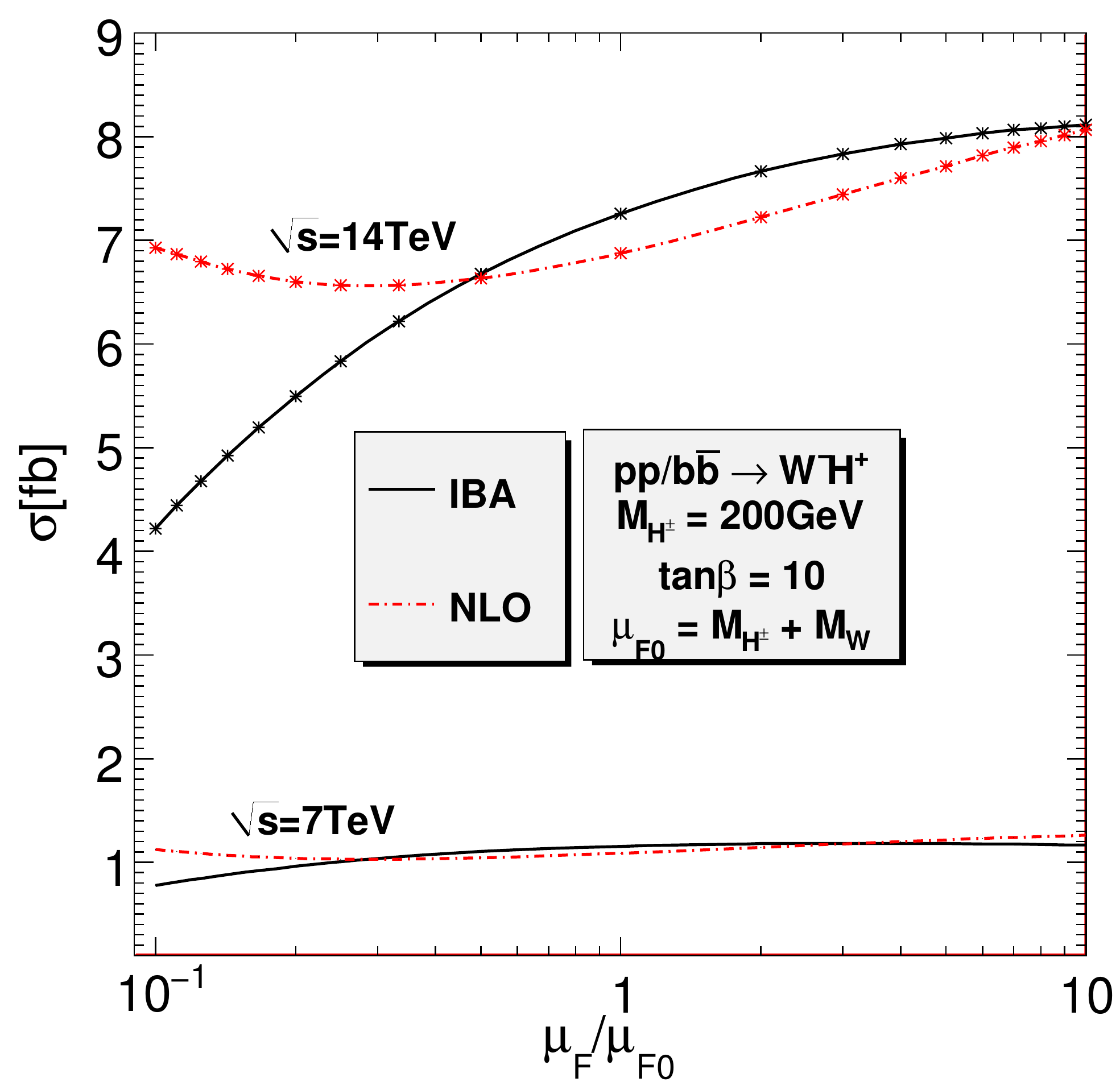}
\hspace*{0.001\textwidth}
\includegraphics[width=0.49\textwidth, height=0.5\textwidth]{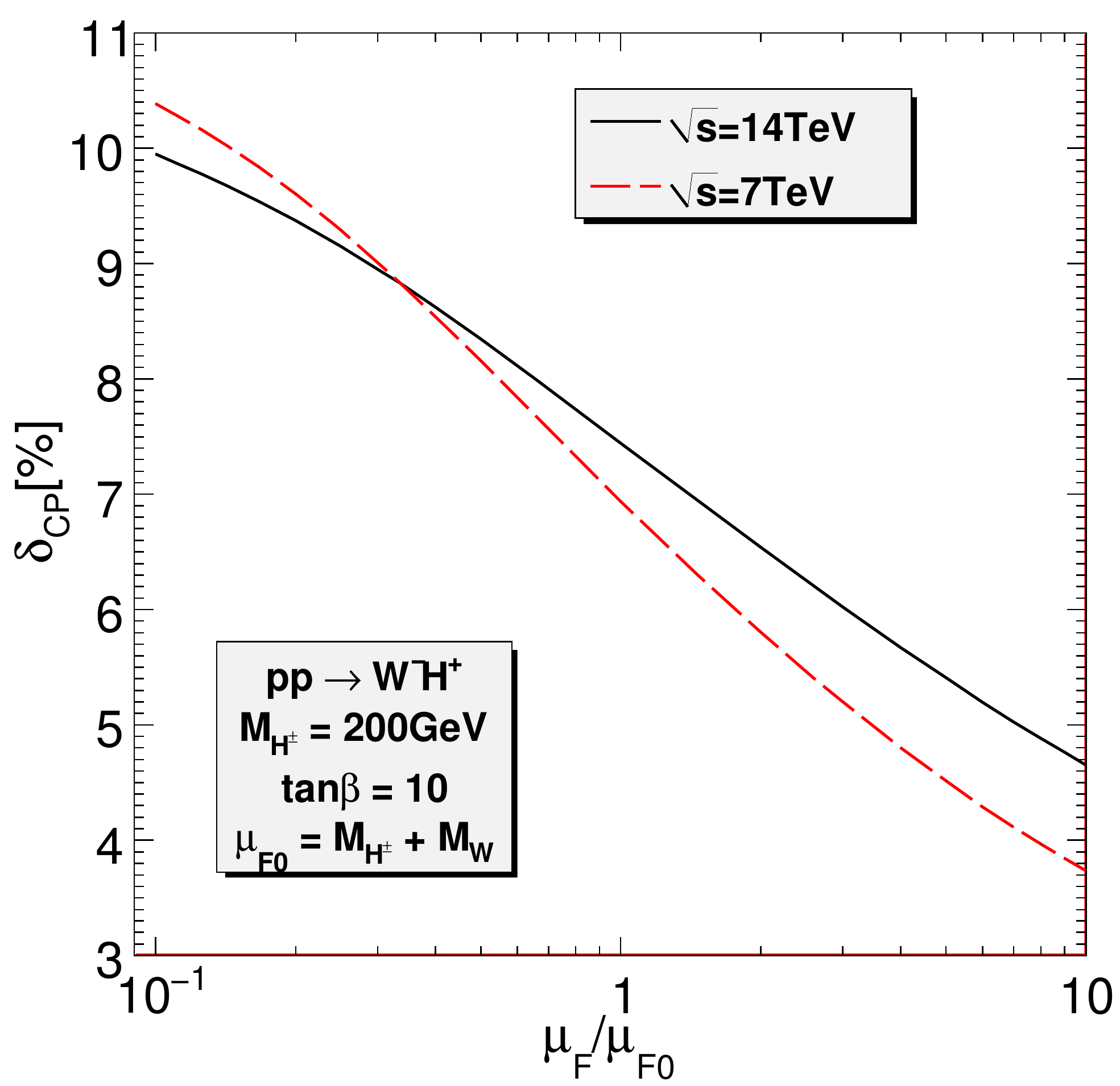}}
\caption{\label{bb_NLO_SCL}{\em The cross section (left) and CP asymmetry (right) as functions 
of the renormalization and factorization scales ($\mu_R=\mu_F$).}}
\end{center}
\end{figure}
In this section we discuss the scale dependence of the total cross sections and CP asymmetries. 
Since the calculation of the loop-induced subprocess $\ggWHpm$ includes only the leading 
order contribution (with improvements on the bottom-Higgs couplings and neutral Higgs-mixing 
propagators), there is no cancellation of the renormalization/factorization-scale dependence 
in this channel. We therefore concentrate on the scale dependence of the
$\bbWHpm$ cross section calculated 
at NLO, see \fig{bb_NLO_SCL} (left). We set $\mu_R = \mu_F$ for simplicity. 
The remaining uncertainty of the NLO scale dependence is 
approximately $9\%$ ($9\%$) when $\mu_F$ is varied between $\mu_{F0}/2$ and 
$2\mu_{F0}$, compared to approximately $14\%$ ($7\%$) for the IBA, at $14\tev$ ($7\tev$) center-of-mass energy. 
The uncertainty is defined as 
$\delta = [\vert\sigma(\mu_{F0}/2)-\sigma(\mu_{F0})\vert + \vert\sigma(2\mu_{F0})-\sigma(\mu_{F0})\vert]/\sigma(\mu_{F0})$. 
The IBA scale dependence looks quite small because we have set 
both renormalization and factorization scales equal, leading to an ``accidental'' cancellation. The IBA cross section increases as 
$\mu_F$ increases while it
decreases as $\mu_R$ increases. We recall that $\mu_F$ enters via the bottom-distribution functions and $\mu_R$ appears in 
the running $b$-quark mass. That accidental cancellation depends strongly on the value of $\tan\beta$. We have verified, by 
studying separately the renormalization and factorization scale dependence, that including NLO corrections does reduce significantly 
each scale dependence. 

\begin{table}[]
 \begin{footnotesize} 
 \bc \caption{\label{table_scale}{ \em Cross sections in$\fb$ 
 for $pp/b\bar{b} \to W^- H^+$ and $pp/gg \to W^- H^+$ at different 
 values of the factorization(renormalization) scale. 
 The CP asymmetries in percentage are also shown.}}
\vspace*{0.5cm}
\begin{tabular}{l r@{.}l r@{.}l r@{.}l r@{.}l r@{.}l r@{.}l r@{.}l r@{.}l}
\hline
 & \multicolumn{8}{c|}{$\sqrt{s}=7\tev$} & \multicolumn{8}{c}{$\sqrt{s}=14\tev$}\\
 \hline
$\mu_R=\mu_F$
&\multicolumn{2}{c}{ $\si_{\text{IBA}}$}
&\multicolumn{2}{c}{ $\si^{b\bar{b}}_{\text{NLO}}$}
&\multicolumn{2}{c}{$\si^{gg}$}
&\multicolumn{2}{c}{$\delta_{\text{CP}}$}
&\multicolumn{2}{c}{ $\si_{\text{IBA}}$}
&\multicolumn{2}{c}{ $\si^{b\bar{b}}_{\text{NLO}}$}
&\multicolumn{2}{c}{$\si^{gg}$}
&\multicolumn{2}{c}{$\delta_{\text{CP}}$}
\\

\hline 
$\mu_{F0}/2$  & 1&1028(2) & 1&0434(3)  & 1&42088(9)  & 8&207(8)   & 6&6774(8) & 6&633(2) &  10&4606(6) & 8&380(7)\\ 
$\mu_{F0}$ & 1&1544(1) & 1&0870(2) & 1&02168(6)  & 6&967(8)  & 7&2568(9) & 6&873(1)&  7&9428(5) & 7&457(8)\\
$2\mu_{F0}$ & 1&1790(1) & 1&1445(2)  & 0&7631(5)  & 5&868(7)  & 7&6648(9) & 7&224(1)&  6&2204(4) & 6&591(8)\\
\hline 
\end{tabular}\ec
 \end{footnotesize}
\end{table}

Concerning the CP asymmetries, the scale dependence is shown in \fig{bb_NLO_SCL} (right). We again set here $\mu_R = \mu_F$. 
If $\mu_F$ is varied between $\mu_{F0}/2$ and 
$2\mu_{F0}$, the uncertainty is approximately $24\%$ ($34\%$) for $14\tev$ ($7\tev$) center-of-mass energy. 
This uncertainty is so large because the dominant contribution to the CP asymmetries (the subprocess $\ggWHpm$) is 
calculated only at LO. 

In \tab{table_scale} we show the values of the cross sections for the two subprocesses as well as the CP asymmetries. 
The scale-dependence uncertainty of the $\ggWHpm$ process is indeed very large. It is mainly due to the running strong 
coupling $\alpha_s(\mu_R)$ which depends logarithmically on the renormalization scale.

\section{Conclusions}
\label{sect-conclusions}
In this paper we have studied the production of charged Higgs bosons in association
with a $W$ gauge boson at the LHC in the context of the complex MSSM. The NLO EW, SM-QCD
and SUSY-QCD contributions to the $b\bar{b}$ annihilation are calculated together with  
the loop-induced $gg$ fusion. 
Special care is dedicated to the use of the effective bottom-Higgs couplings
and the neutral-Higgs boson propagator matrix. 
Moreover, the CP violating asymmetry, dominantly generated by the $gg$ fusion  
parton subprocess, has been investigated. We have
shown that the $\De_b$ and the Higgs-mixing resummations can have large
effects on the production rates and CP asymmetry. 


Numerical results have been presented for the CPX scenario. It is shown
that the production rate and the CP asymmetry depend strongly on $\tan\beta$,
$M_{H^\pm}$ and the phases $\phi_t, \phi_3$. Large production rates
prefer small $\tan\beta$,  small $M_{H^\pm}$ and the phases $\phi_t, \phi_3$
about $\pm \pi$. Large CP asymmetries occur at small $\tan\beta$, 
for $M_{H^\pm}$ of about $250\gev$, and $\phi_t \approx \pm\pi/2$ and $\phi_3 =0$. 

We have also studied the
dependence of the results on the renormalization and factorization scales. 
For the $b\bar{b}$ subprocess, the NLO corrections reduce significantly the scale 
dependence while the $gg$ fusion suffers from large scale uncertainty mainly due to 
the running $\alpha_s(\mu_R)$. This makes the final results, in particular the CP 
asymmetry, depend significantly on the scales. A two-loop calculation would be
needed  to reduce this 
uncertainty to the level of a few percents.\\ 

\noindent{\bf Acknowledgments}\\
We are grateful to Fawzi Boudjema for discussions and for sending us the code $\sloops$.
This work was supported in part by the European Community's Marie-Curie
Research Training Network under contract MRTN-CT-2006-035505
`Tools and Precision Calculations for Physics Discoveries at Colliders'
(HEPTOOLS).

\newpage
\appendix
\section{Feynman diagrams}
We present here the classes of Feynman diagrams that
contribute to the two subprocesses $\bbWHpm$ and $\ggWHpm$. 
We use $h_i$ ($i=1,2,3$) to denote the neutral Higgs bosons ($h,H,A$).

\begin{figure}[h]
  \centering
  \includegraphics[width=0.9\textwidth,height=0.12\textwidth]
{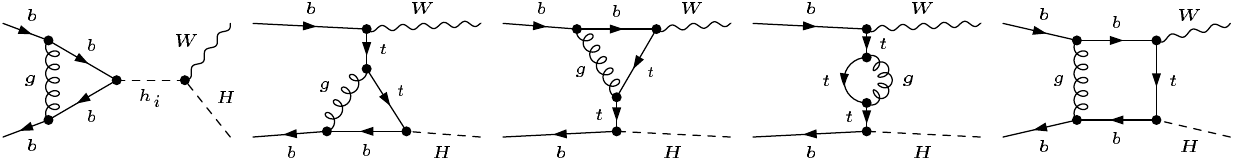}
  \caption{{\em One-loop SM-QCD diagrams for the partonic process 
$\bbWHpm$.}}
  \label{proc_bbWH_SMQCD_virt}
\end{figure}

\begin{figure}[h]
  \centering
 \subfloat[]{ \includegraphics[width=0.8\textwidth,height=0.11\textwidth]
{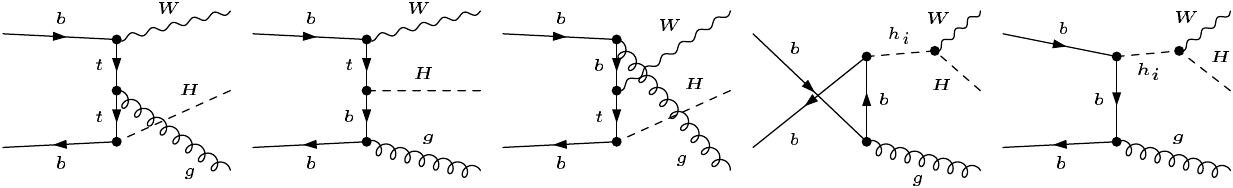}}\\
  \subfloat[ ]{  \includegraphics[width=0.8\textwidth,height=0.11\textwidth]
{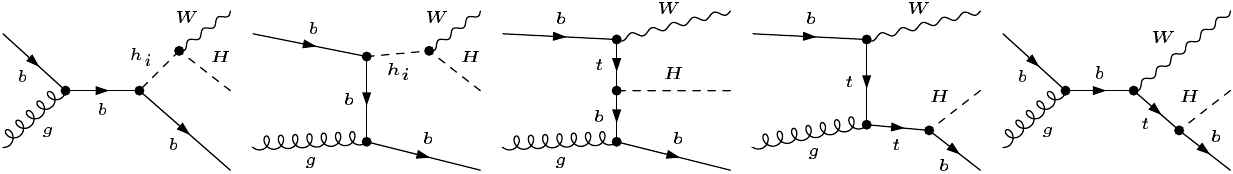}}\\
  \subfloat[ ]{  \includegraphics[width=0.8\textwidth,height=0.11\textwidth]
{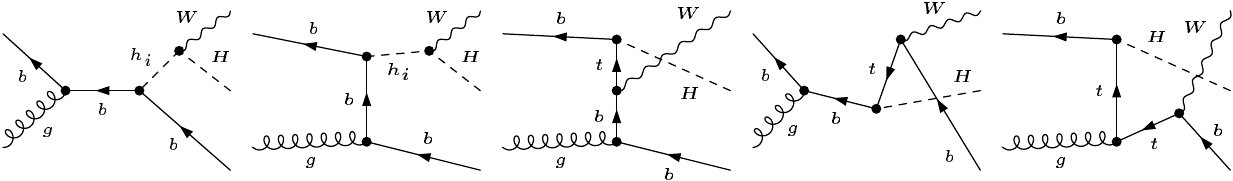}}
  \caption{{\em Gluon-radiation and gluon-induced QCD diagrams.}}
  \label{proc_bbWH_realQCD}
\end{figure}

\begin{figure}
  \centering
  \includegraphics[width=0.9\textwidth,height=0.12\textwidth]
{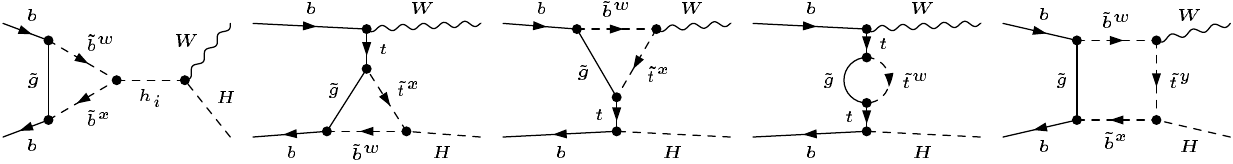}
  \caption{{\em One-loop SUSY QCD diagrams for the partonic process 
$\bbWHpm$.}}
  \label{proc_bbWH_SUSY}
\end{figure}

\begin{figure}
  \centering
  \includegraphics[width=1.\textwidth,height=0.4\textwidth]
{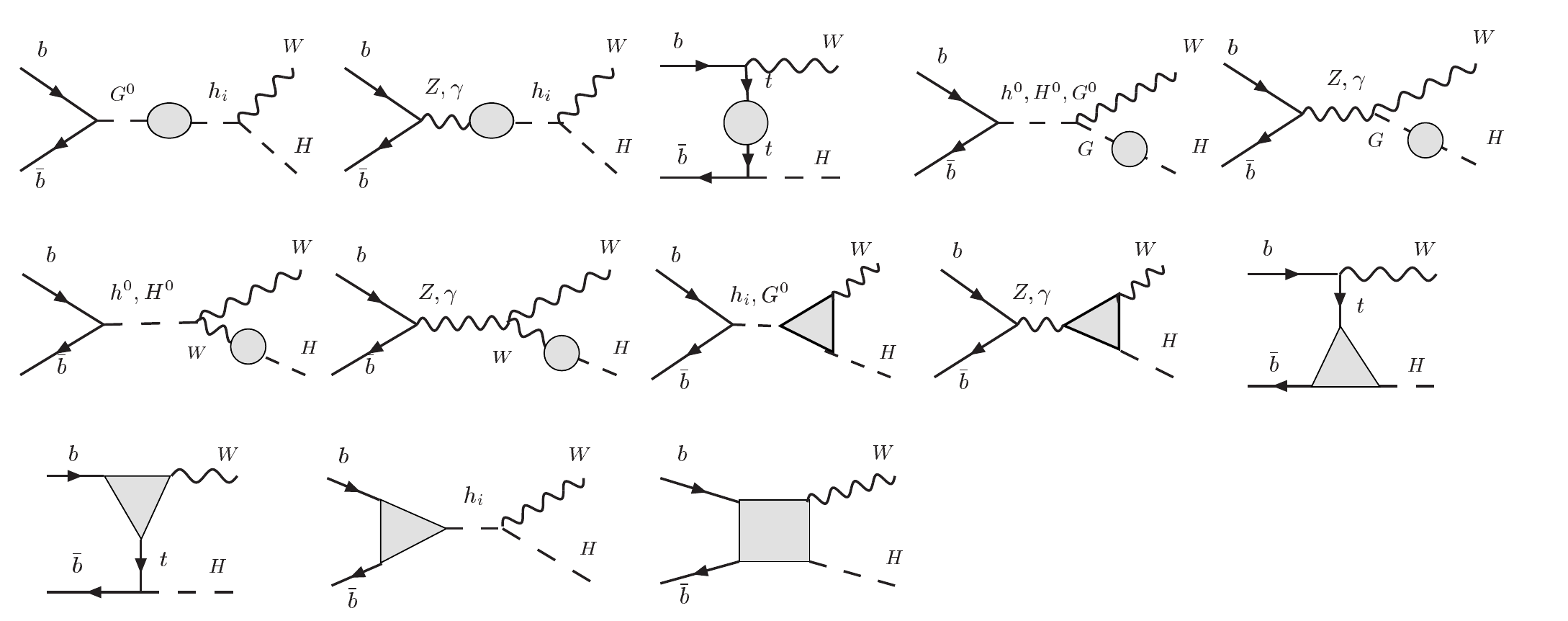}
  \caption{{\em One-loop EW contributions for the partonic process 
$\bbWHpm$. The shaded regions are the one-particle irreducible vertices. 
The diagrams are drawn with Jaxodraw \cite{Binosi:2003yf}.}}
  \label{proc_bbWH_EW}
\end{figure}

\begin{figure}
  \centering
  \includegraphics[width=0.9\textwidth,height=0.6\textwidth]
{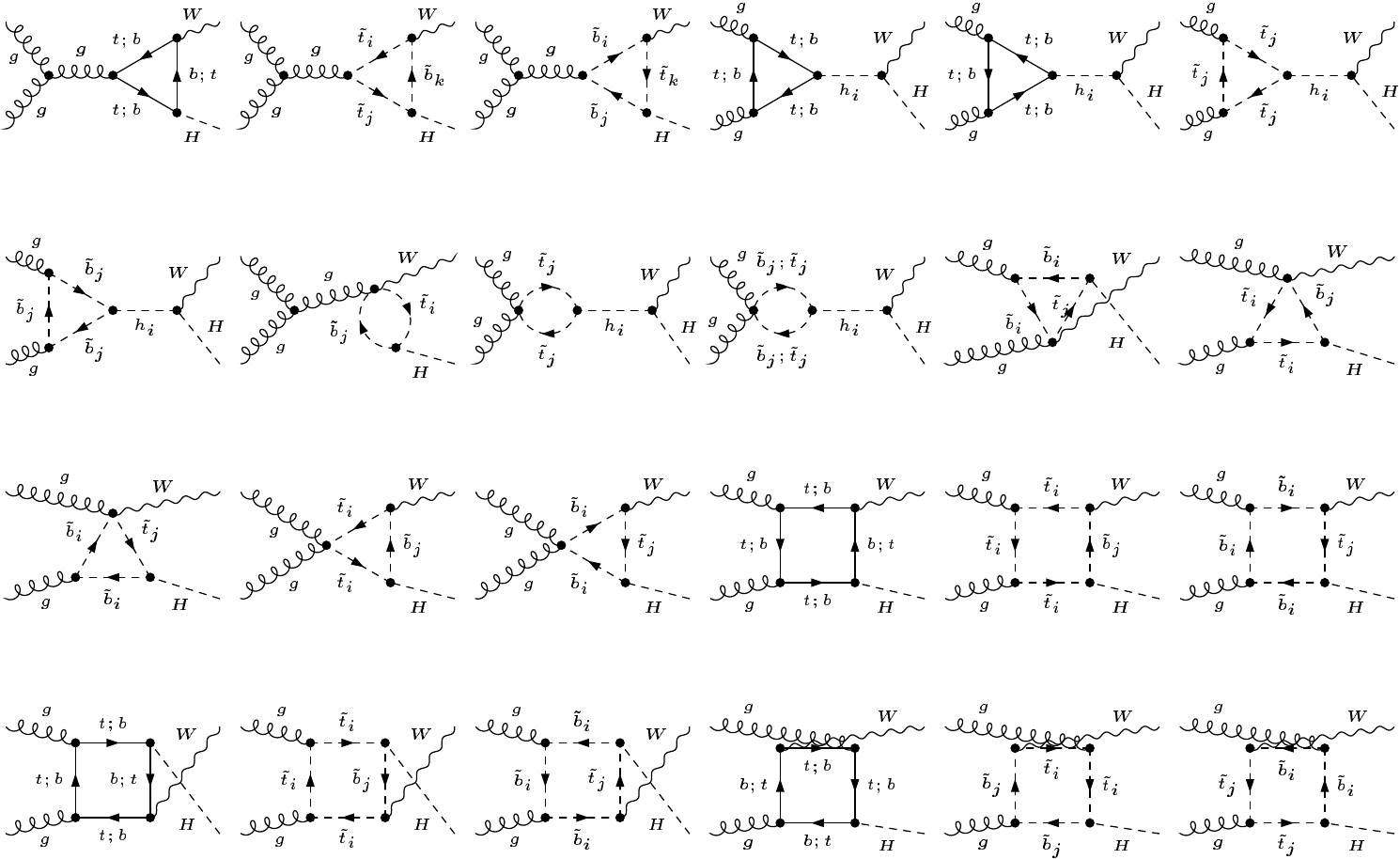}
  \caption{{\em One-loop Feynman diagrams for the partonic process 
$\ggWHpm$.}}
  \label{proc_ggWH}
\end{figure}

\begin{figure}[h]
  \centering
 \subfloat[]{ \includegraphics[width=1\textwidth,height=0.25\textwidth]
{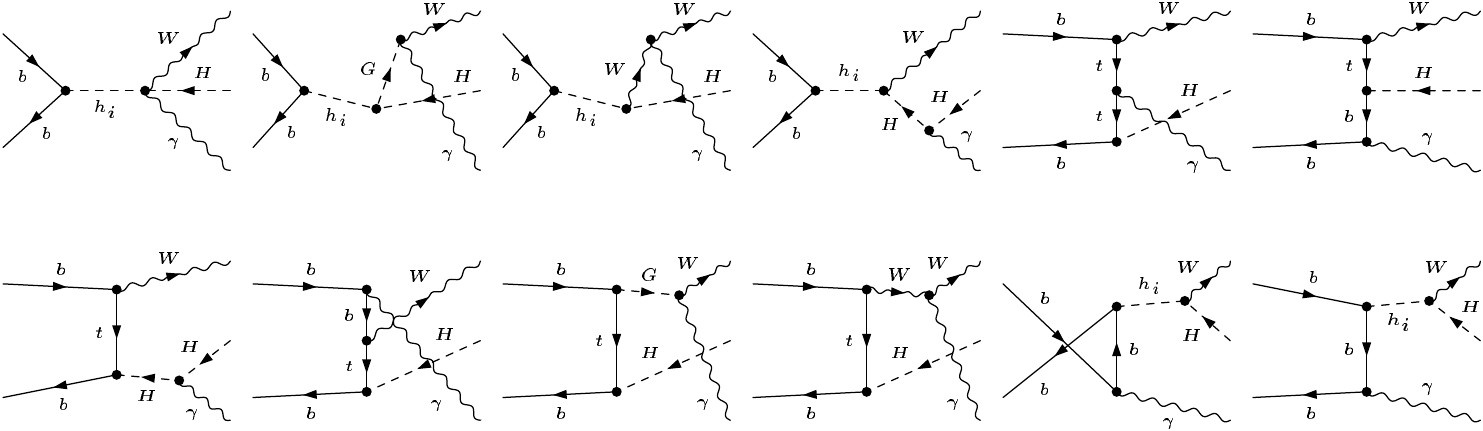}}\\
  \subfloat[ ]{  \includegraphics[width=1\textwidth,height=0.25\textwidth]
{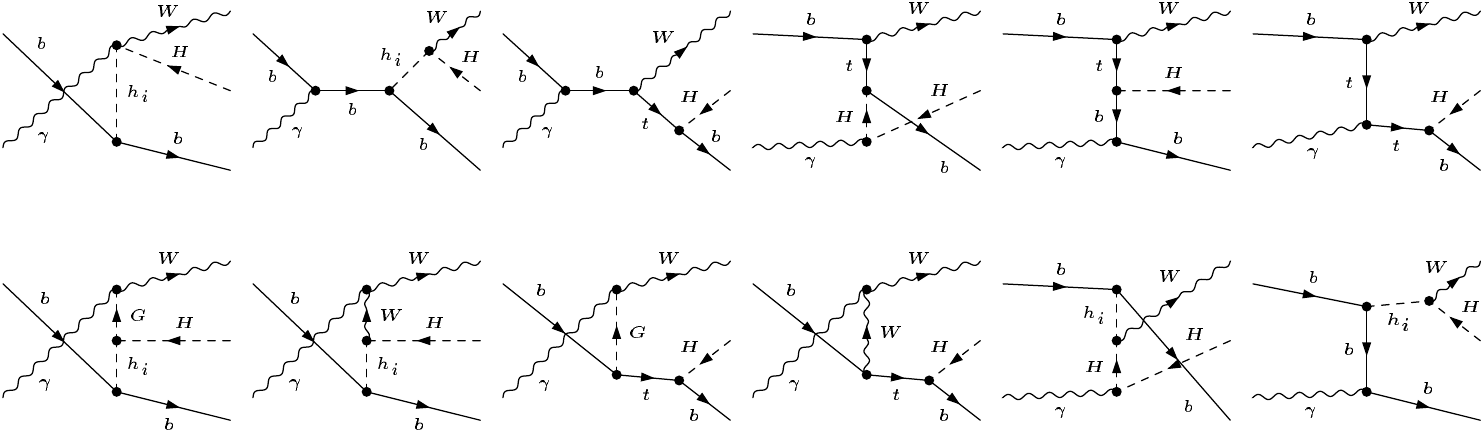}}\\
  \subfloat[ ]{  \includegraphics[width=1\textwidth,height=0.25\textwidth]
{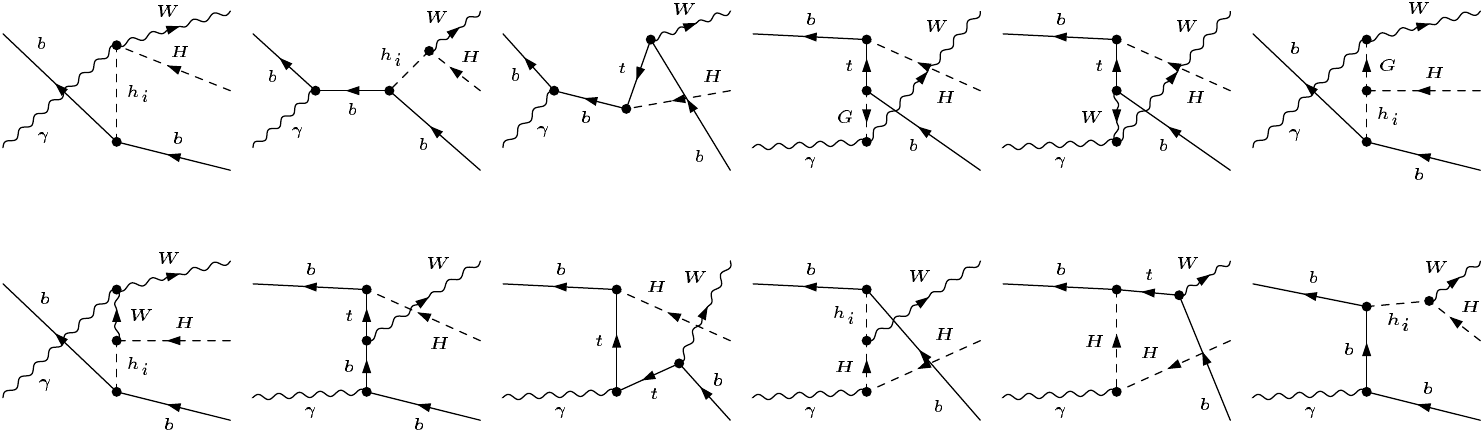}}
  \caption{{\em Photon-radiation and photon-induced EW diagrams.}}
  \label{proc_bbWH_realEW}
\end{figure}

\newpage
\section{Counterterms and renormalization constants}
\label{ap:counterterm}
In this section, we list the Feynman rules and counterterms for vertices and propagators which appear
in the $\bbWHpm$ channel. They can be expressed in terms of coupling
and field renormalization constants (RC) which relate the bare and renormalized quantities.
The RCs are defined as in \bib{Denner:1991kt} for 
the SM-like fields and as in \bib{Frank:2006yh} for the Higgs sector. 
The following one-loop Feynman rules use the standard convention and notation 
of \fas\ \cite{Hahn:2001rv}. In the vertices all momenta are considered as
incoming. We introduce the shorthand notation $s_\alpha = \sin\alpha$, $c_\alpha = \cos\alpha$, $t_\alpha = \tan\alpha$, 
$s_\beta = \sin\beta$, $c_\beta = \cos\beta$, $t_\beta = \tan\beta$.  

\underline{Fermion-Fermion-Scalar:}\\
\begin{figure}[h]
  \centering
  \includegraphics[width=0.4\textwidth]
{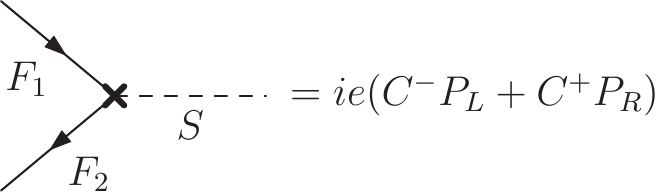}
\end{figure}
\bea \bar b b h^0:&\quad& \begin{cases}
C^- =\fr{  \sa\, m_b}{2\cbb M_Ws_W}\big(1+\de Z_e + \fr{\de m_b}{m_b} +s^2_\beta\de 
\tan\beta- \fr{\de M_W^2}{2M_W^2}-\fr{\de s_W}{s_W}+ \fr12 \de Z_{hh}\crn
 \qquad \qquad \qquad \qquad-\fr1{2\ta}\de Z_{Hh}  + \fr12\de Z_{b,L} + \fr12\de Z_{b,R}^{*} \big) \\
C^+= \fr{  \sa\, m_b}{2\cbb M_Ws_W}\big(1+
\de Z_e + \fr{\de m_b}{m_b} +s^2_\beta\de \tan\beta
- \fr{\de M_W^2}{2M_W^2}-\fr{\de s_W}{s_W}+ \fr12 \de Z_{hh}
\crn \qquad \qquad \qquad \qquad
-\fr1{2\ta} \de Z_{Hh} + \fr12\de Z_{b,L}^{*} + \fr12\de Z_{b,R} \big) \end{cases} \crn
\bar b b H^0:&\quad& \begin{cases}
C^- =-\fr{\ca\, m_b}{2\cbb M_Ws_W}\big(1+
\de Z_e + \fr{\de m_b}{m_b}+s^2_\beta\de \tan\beta- \fr{\de M_W^2}{2M_W^2}-\fr{\de s_W}{s_W}+ \fr12 \de Z_{HH}\crn
 \qquad \qquad \qquad \qquad -\fr12 \ta\de Z_{hH} + \fr12\de Z_{b,L} + \fr12\de Z_{b,R}^{*} \big)\\
C^+= -\fr{\ca\, m_b}{2\cbb M_Ws_W}\big(1+
\de Z_e + \fr{\de m_b}{m_b} +s^2_\beta\de \tan\beta- \fr{\de M_W^2}{2M_W^2}-\fr{\de s_W}{s_W}+ \fr12 \de Z_{HH}
\crn \qquad \qquad \qquad \qquad
 -\fr12 \ta\de Z_{hH} + \fr12\de Z_{b,L}^{*} + \fr12\de Z_{b,R} \big)\end{cases}
\crn
\bar b b A^0:&\quad& \begin{cases}
C^- =-i\fr{  \tbb\, m_b}{2 M_Ws_W}\big(1+
\de Z_e + \fr{\de m_b}{m_b} +s^2_\beta\de \tan\beta- \fr{\de M_W^2}{2M_W^2}-\fr{\de s_W}{s_W}+ \fr12 \de Z_{AA}\crn \qquad \qquad \qquad \qquad
-\fr1{2\tbb} \de Z_{G^0A} + \fr12\de Z_{b,L} + \fr12\de Z_{b,R}^{*} \big)\\
C^+=i\fr{ \tbb\, m_b}{2 M_Ws_W}\big(1+
\de Z_e + \fr{\de m_b}{m_b} +s^2_\beta\de \tan\beta- \fr{\de M_W^2}{2M_W^2}-\fr{\de s_W}{s_W}+ \fr12 \de Z_{AA}\crn \qquad \qquad \qquad \qquad
-\fr1{2\tbb} \de Z_{G^0A} + \fr12\de Z_{b,L}^{*} + \fr12\de Z_{b,R} \big)
\end{cases}
\crn
\bar d_j u_i H^-:&\quad& \begin{cases}
C^- =\fr{\sbb }{\sqrt{2}\cbb s_WM_W}\Big\{
V_{ij}^*\de m_d +V_{ij}^* m_d (1+\de Z_e - \fr{\de s_W}{s_W}+s^2_\beta\de \tan\beta-
\fr{\de M_W^2}{2M_W^2}
\crn \qquad +\fr 12\de Z_{H^-H^+} - \fr12 \de Z_{G^-H^+}/\tbb )
+\fr{m_d}{2} \big[2\de V^*_{ij}+\sum_{k} (V^*_{ik} \de Z^{d*}_{kj,R} + V^*_{kj} \de Z^{u}_{ki,L})\big]\Big\}
\\
C^+= \fr{\cbb}{\sqrt{2}\sbb s_WM_W}\Big\{
V_{ij}^*\de m_u +V_{ij}^* m_u (1+\de Z_e - \fr{\de s_W}{s_W}-c^2_\beta\de \tan\beta-
\fr{\de M_W^2}{2M_W^2} 
\crn \qquad +\fr 12\de Z_{H^-H^+}+ \fr12 \de Z_{G^-H^+}\tbb )
+\fr{m_u}{2} \big[2\de V^*_{ij}+ \sum_{k}(V^*_{ik} \de Z^{d*}_{kj,L} + V^*_{kj} \de Z^{u}_{ki,R})\big] \Big\}

\end{cases}
\crn
\bar u_i d_j  H^+:&\quad& \begin{cases}
C^- =\fr{\cbb}{\sqrt{2}\sbb s_WM_W}\Big\{
V_{ij}\de m_u +V_{ij} m_u (1+\de Z_e - \fr{\de s_W}{s_W}-c^2_\beta\de \tan\beta-
\fr{\de M_W^2}{2M_W^2} 
\crn \qquad +\fr 12\de Z_{H^-H^+}+ \fr12 \de Z_{G^-H^+}\tbb )
+\fr{m_u}{2} \big[2\de V_{ij}+ \sum_{k}(V_{ik} \de Z^{d}_{kj,L} + V_{kj} \de Z^{u*}_{ki,R})\big] \Big\}
\\
C^+= \fr{\sbb }{\sqrt{2}\cbb s_WM_W}\Big\{
V_{ij}\de m_d +V_{ij} m_d (1+\de Z_e - \fr{\de s_W}{s_W}+s^2_\beta\de \tan\beta-
\fr{\de M_W^2}{2M_W^2}
\crn \qquad +\fr 12\de Z_{H^-H^+}  - \fr12 \de Z_{G^-H^+}/\tbb )
+\fr{m_d}{2} \big[2\de V_{ij}+\sum_{k}( V_{ik} \de Z^{d}_{kj,R} + V_{kj} \de Z^{u*}_{ki,L})\big] \Big\}
\end{cases} \\
\label{eq:countertermbbH}\eea

\underline{Fermion-Fermion-Vector:}\\
\begin{figure}[h]
  \centering
  \includegraphics[width=0.4\textwidth]
{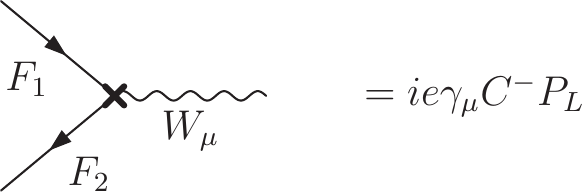}
\end{figure}
\bea \bar d_j u_i W^-:&\quad&  C^- =-\fr{1}{\sqrt{2}s_W}\big[ V^*_{ij} (1+\de Z_e -
\fr{\de s_W}{s_W} + \fr12\de Z_W)+ \de V^*_{ij}
+ \fr12 \sum_{k}(V^*_{kj} \de Z_{ki,L}^{u} +  V^*_{ik} \de Z_{kj,L}^{d*})\big]\crn
\bar u_i d_j W^+:&\quad&  C^- =-\fr{1}{\sqrt{2}s_W}\big[ V_{ij} (1+\de Z_e -
\fr{\de s_W}{s_W} + \fr12\de Z_W)+ \de V_{ij}+ \fr12 \sum_k(V_{kj} \de Z_{ki,L}^{u*} + V_{ik} \de Z_{kj,L}^{d})\big].
\nn\eea

 \underline{Scalar-Scalar-Vector:}\\
\begin{figure}[h]
  \centering
  \includegraphics[width=0.4\textwidth]
{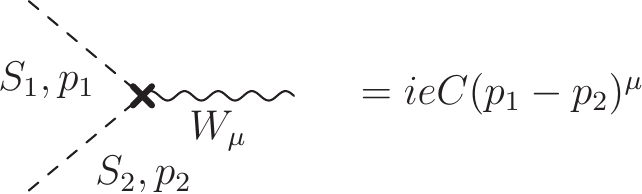}
\end{figure}
\bea hH^-W^+:&\quad& C= -\fr{\cos(\beta-\al)}{2s_W}\big[1+\de Z_e -
\fr{\de s_W}{s_W} + \fr12\de Z_{WW}
                  +\fr 12\de Z_{hh}+\fr 12 \de Z_{H^+H^-}\crn
 &&\qquad \qquad\qquad\qquad-
\fr {\sin(\beta-\al)} {2\cos(\beta-\al)}\big(\de Z_{Hh}
                  -\de Z_{G^-H^+})\big]\crn
hH^+W^-:&\quad& C= \fr{\cos(\beta-\al)}{2s_W}\big[1+\de Z_e -
\fr{\de s_W}{s_W} + \fr12\de Z_{WW}
                  +\fr 12\de Z_{hh}+\fr 12 \de Z_{H^+H^-}\crn
 &&\qquad \qquad\qquad\qquad-
\fr {\sin(\beta-\al)} {2\cos(\beta-\al)}\big(\de Z_{Hh}
                  -\de Z_{G^-H^+})\big]\crn
 HH^-W^+:&\quad& C= \fr{\sin(\beta-\al)}{2s_W}\big[1+\de Z_e -
\fr{\de s_W}{s_W} + \fr12\de Z_{WW}
                  +\fr 12\de Z_{HH}+\fr 12 \de Z_{H^+H^-}\crn
 &&\qquad \qquad\qquad\qquad-
\fr {\cos(\beta-\al)} {2\sin(\beta-\al)}\big(\de Z_{hH}
                  +\de Z_{G^-H^+})\big]\crn
 HH^+W^-:&\quad& C= -\fr{\sin(\beta-\al)}{2s_W}\big[1+\de Z_e -
\fr{\de s_W}{s_W} + \fr12\de Z_{WW}
                  +\fr 12\de Z_{HH}+\fr 12 \de Z_{H^+H^-}\crn
 &&\qquad \qquad\qquad\qquad-
\fr {\cos(\beta-\al)} {2\sin(\beta-\al)}\big(\de Z_{hH}
                  +\de Z_{G^-H^+})\big]\crn
 AH^\pm W^\mp:&\quad& C= -\fr{i}{2s_W}\big[1+\de Z_e -
\fr{\de s_W}{s_W} + \fr12\de Z_{WW}
                  +\fr 12\de Z_{AA}+\fr 12 \de Z_{H^+H^-}\big]\crn
G^0H^\pm W^\mp:&\quad& C= -\fr{i}{4s_W}(\de Z_{AG}+\de Z_{G^-H^+})\crn
\nn\eea
\underline{Vector-Vector-Scalar:}\\
\begin{figure}[h]
  \centering
  \includegraphics[width=0.4\textwidth]
{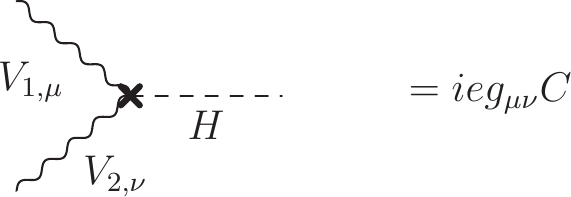}
\end{figure}

The vertices $V W^\mp H^\pm$ with $V=\gamma, Z$ do not appear at tree level. 
The counterterms are generated at one-loop level, however.
\bea \ga W^\pm H^\mp :&\quad&  C= \fr{M_W}{2}(\de Z_{G^-H^+} + \sin{2\beta} \de\tan\beta),\crn
Z W^\pm H^\mp:&\quad& C= -\fr{M_Ws_W}{2c_W}(\de Z_{G^-H^+} + \sin{2\beta} \de\tan\beta).\eea

One also needs counterterms for the renormalized propagators. The complete set of counterterms
for the scalar-scalar case can be found in \bib{Frank:2006yh}. We list here 
extra pieces needed in our calculation.

\underline{Scalar-Vector:}\\
\begin{figure}[h]
  \centering
  \includegraphics[width=0.6\textwidth]
{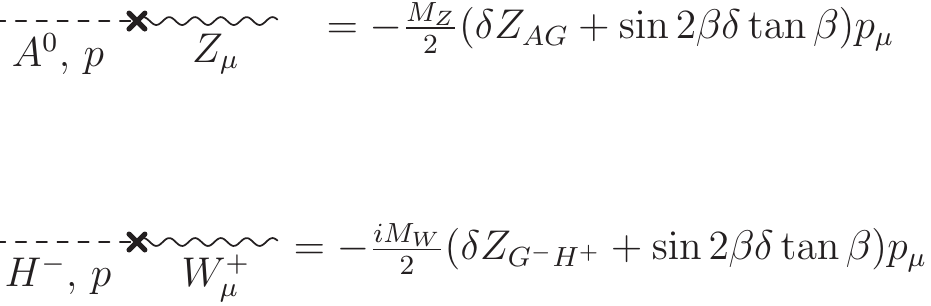}
\end{figure}

\underline{Fermion-Fermion:}\\
The renormalization of the fermion fields in the presence of CP violation is a bit more involved 
than the CP-conserving case. We therefore give here explicit formulae for mass and wave function 
RCs of the quark fields. 
The quark self-energy can be decomposed as
\bea
\Sigma_q(p) = \slashp P_L\Sigma_{q,L}(p^2) + \slashp P_R\Sigma_{q,R}(p^2) + P_L\Sigma_{q,l}(p^2) + P_R\Sigma_{q,r}(p^2).
\eea
We note that $\Sigma_{q,l}(p^2)=\Sigma_{q,r}(p^2)=\Sigma_{q,S}(p^2)$ in the case of CP invariance. 
The renormalized self-energy is written as
\begin{align}
\hat\Sigma_q(p) =& \Sigma_q(p) + \tilde\Sigma_q(p),\\
 \tilde\Sigma_q(p)=& \fr12(\de Z_{q,L}+ \de Z_{q,L}^*)\slashp P_L 
    + \fr12(\de Z_{q,R}+ \de Z_{q,R}^*) \slashp P_L \crn &+ \fr{m_q}2(2\fr{\de m_q}{m_q} 
+ \de Z_{q,L}+ \de Z_{q,R}^*)P_L+\fr{m_q}2(2\fr{\de m_q}{m_q} 
+ \de Z_{q,L}^*+ \de Z_{q,R})P_R.
\end{align}
 In general, $\delta Z_{q,L}$ and $\delta Z_{q,R}$ are 
complex\footnote{If we impose CP invariance then $\delta Z_{q,L}$ and $\delta Z_{q,R}$ can be taken real.}. $\delta m_q$ 
can always be made real by rephasing the field $\psi_L$ (or $\psi_R$). At this step any phases can be absorbed into 
the two factors $\delta Z_{q,L}$ and $\delta Z_{q,R}$ which will have to be determined. It is obvious that the squared 
amplitude is invariant under a global rephasing
\bea
\psi = \psi_L + \psi_R \longrightarrow  e^{i\phi_\psi}(\psi_L + \psi_R ).
\label{phase_q_free}
\eea    
From this freedom we can, for example, make $\delta Z_{q,R}$ real while $\delta Z_{q,L}$ remains complex. We therefore 
need four conditions to determine the three renormalisation constants. 
The OS conditions are
\bea
\REt\hat\Gamma_q(p)u(m_q)&=&0, \hs \left[\fr{1}{\slashp-m_q}\REt\hat\Gamma_q(p)\right]u(m_q)=iu(m_q),\label{Sigma_q_cd1}\\
\REt\bar{u}(m_q)\hat\Gamma_q(p)&=&0, \hs \bar{u}(m_q)\left[\REt\hat\Gamma_q(p)\fr{1}{\slashp-m_q}\right]=i\bar{u}(m_q),\label{Sigma_q_cd2}
\eea
where $\hat\Gamma_q(p)=i[\slashp - m_q + \hat\Sigma_q(p)]$ and $p^2=m_q^2$. 
$\REt$ sets the imaginary part of the loop integrals to zero since they are not 
involved in the renormalisation. 
The Hermiticity of the Lagrangian imposes\cite{Aoki:1982ed}
\bea
\hat\Gamma_q(p) = -\gamma_0^\dagger \hat\Gamma_q^\dagger (p)\gamma_0.
\label{Gamma_q_hermiticity}
\eea
It is obvious that \eq{Sigma_q_cd2} can be derived from \eq{Sigma_q_cd1} and \eq{Gamma_q_hermiticity}. From these conditions we 
get the following results
\bea
\delta m_q &=& \fr{1}{2}\REt\Big\{m_q\big[\Sigma_{q,L}(m_q^2) + \Sigma_{q,R}(m_q^2) \big] + \Sigma_{q,l}(m_q^2) + \Sigma_{q,r}(m_q^2) \Big\},\crn
\delta Z_{q,L} &=& -\REt\Big\{\Sigma_{q,L}(m_q^2) - \fr{1}{2m_q}\big(\Sigma_{q,l}(m_q^2) - \Sigma_{q,r}(m_q^2)\big) \crn
&+& m_q\fr{d}{dp^2}\big[m_q\big(\Sigma_{q,L}(p^2) + \Sigma_{q,R}(p^2)\big) 
+ \Sigma_{q,l}(p^2) + \Sigma_{q,r}(p^2) \big]_{p^2=m_q^2} \Big\}, \crn
\delta Z_{q,R} &=& -\REt\Big\{\Sigma_{q,R}(m_q^2) - \fr{1}{2m_q}\big(\Sigma_{q,r}(m_q^2) - \Sigma_{q,l}(m_q^2)\big) \crn
&+& m_q\fr{d}{dp^2}\big[m_q\big(\Sigma_{q,L}(p^2) + \Sigma_{q,R}(p^2)\big) 
+ \Sigma_{q,l}(p^2) + \Sigma_{q,r}(p^2) \big]_{p^2=m_q^2} \Big\},
\eea
where we have used the freedom \eq{phase_q_free} to take $\IM(\delta Z_{q,R})= -\IM(\delta Z_{q,L})$. These results agree with 
the ones in \cite{Williams:2008phd}. 


\end{document}